\documentclass{article}
\usepackage{graphicx} % Required for inserting images
\usepackage{amsmath}
\usepackage{amsfonts}
\usepackage{float}
\usepackage{url}
\usepackage[toc,page]{appendix}
\usepackage{tabularx}
\usepackage[margin=1in]{geometry}
\usepackage[normalem]{ulem}
\usepackage{xcolor}
\usepackage{booktabs}
\usepackage{listings}
\usepackage{pdflscape}
\usepackage[T1]{fontenc}
\lstdefinestyle{Rstyle}{
  language=R,
  basicstyle=\ttfamily\small,
  upquote=true,
  columns=fullflexible,
  keepspaces=true,
  showstringspaces=false,
  breaklines=true,
  commentstyle=\itshape\color{gray!60!black}
}
\newcommand{\code}[1]{\texttt{#1}}

\title{Reservoir: A Large-Scale Simulated Dataset for Training and Evaluating Epidemiological Models}
\date{\today}
\date{\today}

\author{
Carson Dudley$^{1, 2, *}$ \and
Reiden Magdaleno$^{2,*}$ \and
Marisa Eisenberg$^{1, 2, 3}$
}

\begin{document}

\maketitle

\begin{center}
{\small
$^1$ Department of Mathematics, University of Michigan\\
$^2$ School of Public Health, University of Michigan\\
$^3$ Center for the Study of Complex Systems, University of Michigan\\[4pt]
$^*$ Equal contribution\\[4pt]
{\small \texttt{\{cdud, reiden, marisae\}@umich.edu}}}
\end{center}

\begin{abstract}

Large-scale, standardized datasets have driven many advances in AI-based scientific modeling, from protein structure prediction to natural language processing. Infectious disease epidemiology is increasingly adopting AI methods for forecasting, surveillance, and outbreak analytics, but the time-series data available to train them remains orders of magnitude smaller than the corpora behind the advances seen in other fields. Because the scope of real-world epidemiological data cannot practically reach the scale needed to train truly large-scale AI methods, simulated data provides a possible alternative. Here we introduce Reservoir, a large open simulator and dataset of realistic epidemic simulations in which every trajectory carries complete ground-truth labels, including quantities that cannot be measured directly in a real outbreak, such as true infection counts, time-varying reproduction numbers, and counterfactual intervention effects. Reservoir is generated by a stochastic simulator with realistic noise and reporting artifacts, together with interventions with configurable timing, compliance, and age-dependent efficacy. The current release contains 500,000 outbreak trajectories spanning one billion simulated days across diverse pathogen characteristics, population structures, and intervention regimes. Reservoir enables counterfactual experiments, surveillance-design studies, and training of epidemic models at a scale real-world datasets cannot provide.

\end{abstract}

\section*{Motivation and significance}

\noindent Artificial intelligence has transformed scientific domains where large datasets exist. In structural biology, AlphaFold---a deep learning system for predicting three-dimensional protein structure from amino acid sequences, which won the 2024 Nobel Prize in Chemistry---was enabled by compiling decades of curated protein structures in the Protein Data Bank \cite{alphafold, pdb}. Large language models trained on massive text corpora have reshaped natural language processing, with scaling laws demonstrating that performance improves predictably with data volume \cite{gpt2, gpt3, scalinglaws}. Across these and other domains, there is a consistent pattern: dataset scale is a primary driver of model capability.

Infectious disease epidemiology is increasingly turning to AI methods. Neural networks and other machine learning approaches are now used for outbreak forecasting \cite{deepgleam}, surveillance signal detection \cite{gao2025early}, parameter inference \cite{sgnns2025}, and intervention evaluation \cite{mu2025counterfactualprobabilisticdiffusionexpert}. However, the data available to train these models is several orders of magnitude smaller than what has driven progress in other fields. Real-world surveillance datasets, while extremely valuable, often only cover a single country or a comparatively narrow range of pathogens or surveillance modalities versus what is possible \cite{project_tycho, delphi_epidata}, and small population data or aggregation across jurisdictions is often necessarily constrained by data-sharing agreements and privacy protections \cite{pew2024publichealth}. Real world data also cannot provide ground truth for latent quantities such as true infection counts, time-varying reproduction numbers, or counterfactual intervention effects, which cannot typically be directly measured in real outbreaks.

To address this gap, we developed Reservoir: a large open-source simulator and dataset of realistic epidemic simulations. The simulator contains two parts: a scientifically valid disease simulator, and a realistic observation model, which takes the true underlying infectious disease dynamics and adds real-world effects like underreporting, reporting delays, and overdispersion to the clean simulations. Each simulation records both the underlying epidemic trajectory---true infections, transmission rates, reproduction numbers, and intervention effects---and the noisy observations that a public health agency would actually see, enabling direct comparison of estimated quantities against known truth. Reservoir samples broadly across pathogen biology, demographic structure, surveillance regimes, and intervention policies, producing trajectories that span the range of dynamics observed in real-world data. The current release contains 500,000 outbreak trajectories spanning one billion simulated days. Reservoir was used to train Mantis, a foundation model for infectious disease forecasting that achieved high accuracy out-of-the-box across 16 real-world diseases despite only being trained on simulated data \cite{mantis}.

Beyond serving as training data for AI models, Reservoir supports a range of uses that real-world data practically cannot. Because ground truth is available, established methods (e.g. for calculation of quantities such as serial interval, reproduction numbers, etc.) can be benchmarked against known answers under controlled and progressively challenging conditions. In addition to benchmarking, this enables applications in parameter inference \cite{ganyani2020estimating}, intervention effect estimation \cite{barros2022causal}, inference robustness assessment \cite{koopman2004modeling}, and other tasks where validation has historically relied on simplified simulations or expert judgment. The configurability of the simulator also enables counterfactual experiments, surveillance system design studies, and systematic investigation of how data quality affects inference. The complete dataset and simulator code are publicly available at \texttt{https://github.com/micom-hub/Reservoir}. 

\section*{Software Description}
% \section{Simulator Design}

\noindent The Reservoir package is a stochastic epidemic simulator implemented in C++ and R. It is designed to generate large, diverse synthetic datasets of epidemic trajectories that can be used for benchmarking, methods development, and AI model training. To successfully install Reservoir, a C++ compiler must be downloaded and installed first. Reservoir can then be installed from Github using the following command (using the \verb|devtools| package or a similar packages):

\begin{lstlisting}[style=Rstyle]
> devtools::install_github("micom-hub/Reservoir")
> library(Reservoir)
\end{lstlisting}

%Reservoir consists of 3 C++ scripts and R 11 scripts. % Seem to be more files in the github, plus not sure we need this
This article will focus on the main dataset generation and parameter defining functions: \texttt{reservoir\_config()}, \texttt{model\_structure()}, \texttt{population\_spec()}, \texttt{intervention\_policy()} and
\texttt{generate\_dataset()}. The package also has an accompanying Github page with additional vignettes and tutorials (e.g. see the "getting started" tutorial on the Github page linked above). % the url is on the page already, maybe we can ditch this?

\section*{Software Architecture}
\noindent The purpose of Reservoir is to generate realistic epidemic scenarios that span a wide range of diseases, reporting effects, surveillance systems, population structures, and public health interventions. A visual summary of the overall Reservoir workflow is shown in Figure~\ref{fig:conceptual}. Reservoir supports three transmission mechanisms: human-to-human, vector-borne, and environmental. 
Stochastic epidemic simulations are generated using a fast C++ adaptive tau-leaping engine coupled with a hierarchical R parameter sampler and an observation model that maps true incidence onto surveillance signals.
Before the adaptive tau-leaping engine is executed, the sampled parameters and model structure are assembled into configuration objects in R. Users can set initial conditions, define compartments and transition rates, and  population structures.
The user can define a single or multiple transmission mechanisms that subsequently call the appropriate sampler and C++ simulator, using \texttt{generate\_dataset()} to run them in a single batch. Detailed documentation for each function is available in the GitHub repository accompanying this paper.

 \begin{figure}
    \centering
     \includegraphics[width=\textwidth,  height=\textheight, keepaspectratio]{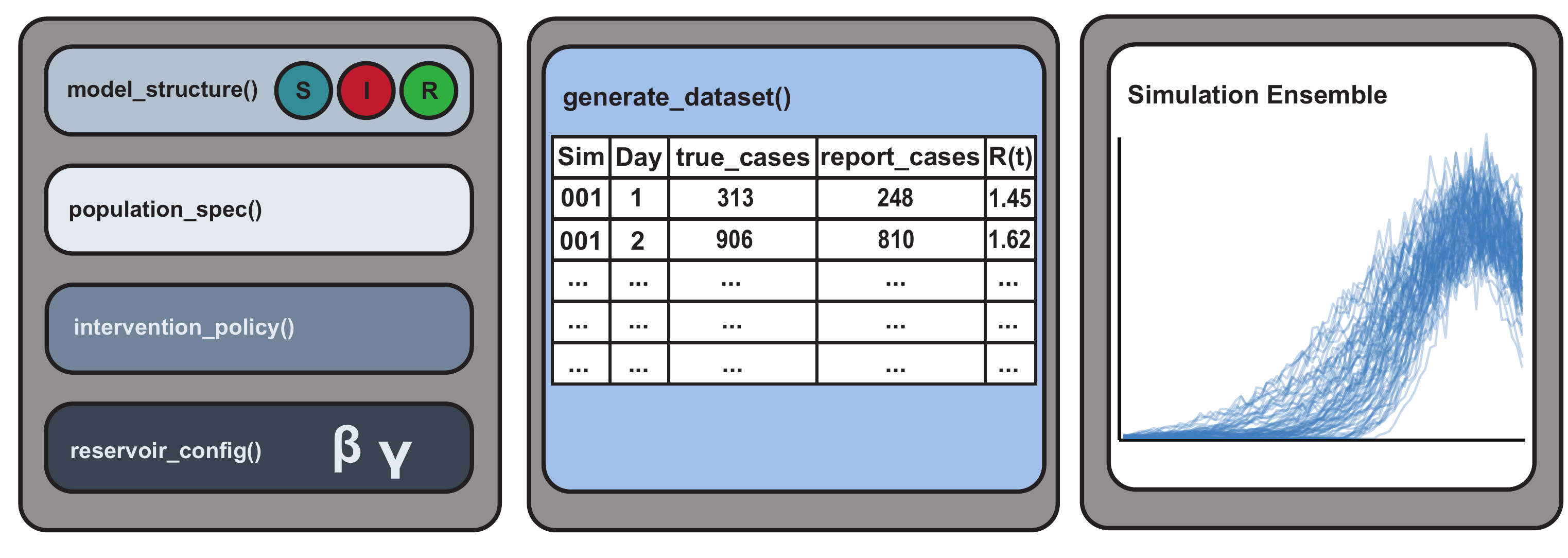}
 \caption{\textbf{Dataset Generation Workflow.} To generate a synthetic epidemic dataset the user will first define the model structure, population and initial conditions, intervention policy, and disease parameter values using the following functions: \texttt{reservoir\_config()}, \texttt{model\_structure()}, \texttt{population\_spec()}, and \texttt{intervention\_policy()}. The user then calls \texttt{generate\_dataset()} to produce the synthetic disease dataset that contains both the true underlying epidemic dynamics and noisy observations in real public health data. With these functions, the user can build a single disease simulation or an ensemble of disease trajectories for tasks that include AI model training, parameter estimation, and intervention evaluations.
    }
    \label{fig:conceptual}
\end{figure}

\section*{Software Functionalities}

\noindent A default-settings dataset can be generated in a single function call (using \verb|generate_dataset()|). For a user to generate a dataset with specified initial conditions and parameters, they need the five R functions outlined below. The pipeline is structured such that four setup functions first define what is being simulated (compartmental structure, population, intervention schedule, prior distributions). These setup functions are then passed to \texttt{generate\_dataset()}, which runs the simulation and returns the sampled parameters, true incidence, daily \texttt{$R_t$}, compartment counts, and simulated surveillance signals. More details can be found in the SI and tutorials on Github.
\medskip

\noindent\textbf{Model structure (\texttt{model\_structure()}).}
The \texttt{model\_structure()} function allows the user to initialize the compartment structure and respective transitions through eleven arguments: (\texttt{has\_latent}) adds a latent stage to the transmission model, (\texttt{has\_asymptomatic}) includes an asymptomatic infectious class,  and  (\texttt{has\_recovery}) determines the existence of a recovered state. Recovered individuals can have different levels of immunity through (\texttt{immunity}) that include \texttt{"waning"}, \texttt{"permanent"}, or \texttt{"none"}. Waning immunity can also occur through one, two, or three sequential compartments by defining (\texttt{n\_waning\_stages}) where the duration of immunity can be Gamma-distributed.

Intervention compartments can also be added through  the (\texttt{has\_isolation}), (\texttt{has\_quarantine}), (\texttt{has\_pep}), and (\texttt{has\_prep}) arguments. Users can also specify population demographics (\texttt{has\_demography}) and case importation (\texttt{has\_importation}). Three different mechanisms can also be determined using the (\texttt{mechanisms}) argument and calling one or a mix of the following: \texttt{"base"}, \texttt{"vectorborne"}, and \texttt{"environmental"}. 

\medskip

\noindent\textbf{Population and initial conditions (\texttt{population\_spec()}).}
The population patches and initial conditions are called together through \texttt{population\_spec()}, where population is divided into  $P$ spatial patches and $A$ age strata (\texttt{num\_pops}, \texttt{num\_ages}), yielding different  $P \times A$ subpopulations. By default, the total population size (\texttt{pop\_size}) spans five thousand to forty million individuals and the simulation horizon (\texttt{num\_days}) set as five thousand days. Contact and mixing matrices can be explicitly provided by the user (\texttt{contact\_matrix}, \texttt{mixing\_matrix}) or generated automatically by the hierarchal sampler.

Initial conditions can be specified in two ways: a proportion of the population can be initialized as infectious using  (\texttt{seed\_frac}) or explicit initial infection counts can be drawn from a distribution $\sim\!\mathcal{U}(1,...,50)$ with (\texttt{n\_seed}).  In the explicit form, the index cases are placed into a single compartment chosen from $L$, $I_s$, and $I_a$ at varying probabilities. The fractional form has an additional accompanying argument that declares whether the pathogen is novel (\texttt{novel}). When it is not, a fraction of the population starts immune from (\texttt{immune\_frac}) or derived from the endemic equilibrium $1 - 1/R_0$. More details regarding initialization can be found in the appendix.

\medskip

\noindent\textbf{Intervention policy (\texttt{intervention\_policy()}).}
Interventions can be implemented according to a fixed schedule or initiated after an epidemic threshold is reached in a simulation by calling \texttt{intervention\_policy()}. Under the scheduled approach, interventions begin on a specified day (\texttt{start\_day}) and remain active for a defined duration  (\texttt{duration}). In the reactive mode, interventions activate after the true incidence exceeds an upper threshold  (\texttt{on\_threshold}) with a detection lag applied  (\texttt{trigger\_delay}). Interventions end after incidence falls below  a lower threshold (\texttt{off\_threshold}). Thresholds can either be written as counts when values are greater than or equal to one or can be written as fractions when counts are less than 1.

Intervention measures occur through two mechanisms. At the compartment level, interventions are parameterized through daily per-capita rates: (\texttt{iso\_rate}) controls the movement of infectious individuals into isolation, (\texttt{quar\_rate}) the quarantining of susceptible and latent individuals, (\texttt{pep\_rate}) the administration of post-exposure prophylaxis, and (\texttt{prep\_start\_rate}) the rate of preexposure prophylaxis with (\texttt{prep\_eff}) defining the efficacy. Three arguments school closure (\texttt{school\_closure}), workplace closure (\texttt{work\_closure}), and reactive contact reduction (\texttt{contact\_reduction}) act on the transmission process itself, changing the contact structure of the whole population as opposed to individual states. The probability that an individual simulation has intervention effects is defined by (\texttt{enable\_prob}).

\medskip

\noindent\textbf{Prior distributions and observation model (\texttt{reservoir\_config()}).}
All distributional assumptions are collected in a single object through \texttt{reservoir\_config()}. Arguments can either only accept a single value, or both a single value and a pair of values. Defining a single value fixes the parameter for the simulation, whereas defining a pair of values \texttt{c(lower, upper)} sets the the central 90\% of the sampling distribution from which the parameter is drawn from. The main epidemiological quantities are the following and can accept both a single value or pair: mechanism-specific basic reproduction number  (\texttt{r0\_base}, \texttt{r0\_vector}, \texttt{r0\_waterborne}),  the infectious and latent periods (\texttt{infectious\_days}, \texttt{latent\_days}), and the duration of immunity   (\texttt{immunity\_days}).  Other arguments characterize additional transmission dynamics like super-spreading intensity, case importation, seasonal forcing, the timing and magnitude of transmission waves, and the asymptomatic transmissibility and fraction. Additionally, arguments for parameters related to demographic rates and intervention effectiveness are also drawn. 

The observation-model parameters are also held in the same object within \texttt{reservoir\_config()}. This includes parameters for case ascertainment (underreporting), reporting delay, negative-binomial overdispersion (day-to-day variation), hospitalization and case-fatality probabilities, and the probability with which each reporting artifact is applied to true incidence. Users can specify which signals are generated since the $R_t$ calculation and the full observation model dominate runtime for larger datasets.  With the argument \texttt{scope}, users can select \texttt{"full"} (all observation signals  with $R_t$) , \texttt{"signals"} (all observation signals but no $R_t$), or \texttt{"cases\_only"} (true incidence only). They can also individually select which signals to keep by declaring \texttt{TRUE} or \texttt{FALSE} across the following five arguments: \texttt{compute\_rt} (the daily $R_t$ series), \texttt{compute\_noise\_full} (the daily $R_t$ series), \texttt{compute\_wastewater} (wastewater signal), \texttt{compute\_wastewater} (wastewater signal), \texttt{compute\_syndromic} (syndromic consultation signal), and \texttt{compute\_hosp\_deaths} (hospitalizations, deaths, and occupancy). Collectively, these observational effects transform the original model simulations into more realistic imperfect data similar to those observed in real-world disease surveillance systems.  For more details on the parameters used in each real world effect model, please see the Appendix or Github tutorial.

\medskip
\noindent\textbf{Dataset generation (\texttt{generate\_dataset()}).}
The simulation pipeline can be ran with a single call using \texttt{generate\_dataset()}. One or more transmission mechanisms can be defined with \texttt{mechanisms} with \texttt{n\_per\_mechanism} simulation runs collected for each. This function takes the \texttt{model\_structure()}, \texttt{population\_spec()}, \texttt{intervention\_policy()}, and \texttt{reservoir\_config()} configurations described in earlier sections as arguments and  returns a list of simulations. Each run is processed through hierarchical parameter sampling, configuration, initial-condition specification, stochastic simulation, $R_t$ calculation, and the observation-model application. By default, simulation runs in which the epidemic dies out immediately are discarded and reran. The \texttt{n\_per\_mechanism} counts successful runs and redrawing stops after \texttt{max\_attempts}.

The returned list carries the following elements:  sampled parameters, the core transmission model structure, the true daily incidence, daily $R_t$, compartment counts, and simulated surveillance signals: reported cases, hospitalizations, deaths, hospital occupancy, wastewater concentration, and a syndromic consultation signal.  An additional function \texttt{extract\_all\_runs\_single\_csv()} can write saved datasets to a long-format table, where the structural labels (\texttt{has\_latent}, \texttt{has\_pep}, etc.) alongside time-varying fields (compartments, signals, daily $R_t$)are columns and run-days are rows .

\section*{Illustrative Example}
\noindent In this section, we provide two illustrative examples of generating simulated data using Reservoir. The first example generates an ensemble of epidemic outbreaks only using \texttt{generate\_dataset()} and the second example shows how users can specify their own model structure, disease parameters, and initial conditions. These and several additional examples are also provided in the respective Github repository. 

\subsection*{Example 1: Dataset generation}
The following example uses the Reservoir R package to generate 10000 synthetic epidemic outbreaks of an environmentally transmitted pathogen. It demonstrates how a large-scale labeled training dataset can be produced. Likewise, the example also shows how the observed surveillance signals appear compared to the true epidemic dynamics. The R package is loaded at the start of the session with  \texttt{library()} and subsequently compiles the C++ engines.

\begin{lstlisting}[style=Rstyle]
library("Reservoir")

runs <- generate_dataset(
  n_per_mechanism = = 10000,
  mechanisms = "environmental",
  save_dir = "sim_data",
  seed = 1
)
\end{lstlisting}

This example uses the \texttt{generate\_dataset()} function and arguments that are not specified are set to default. Defining the argument \texttt{seed} makes simulation runs reproducible and passing a directory path through \texttt{save\_dir()} writes an \texttt{<mechanism>\_runs.rds} file in that respective directory. Although not used in this example, the configuration arguments for the model structure, population geometry, intervention policy, and prior distributions are set with \texttt{model\_structure()}, \texttt{population\_spec()}, \texttt{intervention\_policy()}, and \texttt{reservoir\_config()} respectively. 

A single call runs the following pipeline for each simulation: hierarchical parameter sampling, configuration building, initial condition, stochastic simulation, $R_t$ computation, and application of the observation model. Because 
\texttt{n\_per\_mechanism} only counts successful epidemic simulations rather than attempts, the default distribution of parameters across a generated dataset is not the prior it was drawn from. The distribution of parameters is now the prior conditioned on the epidemic taking off. Draws with $R_0$ close to one, low initial infection counts, or short infectious periods are more likely to be discarded. A run is retained if it produces at least five infections in endemic disease scenarios or ten otherwise. This rule can adjusted by changing the parameters for the \texttt{min\_cases\_endemic} and \texttt{min\_cases\_nonendemic} arguments in \texttt{reservoir\_config()}.

 \begin{figure}
    \centering
     \includegraphics[width=\textwidth,  height=\textheight, keepaspectratio]{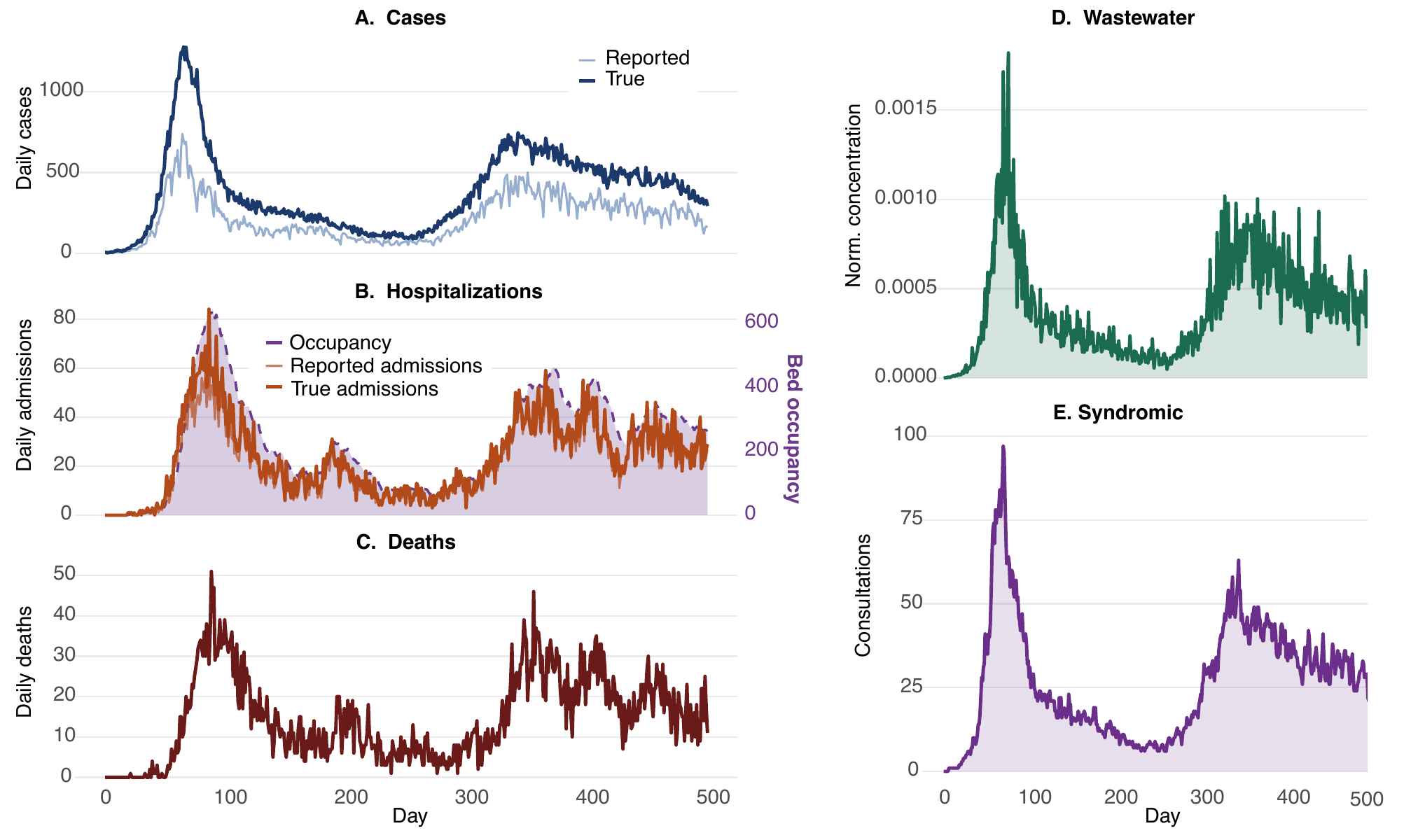}
    \caption{\textbf{Example trajectory for one Reservoir simulation.} (Top left) Reservoir reports both true and reported cases for each outbreak trajectory. (Middle left) Reservoir reports true and reported hospital admissions, as well as hospital occupancy as another surveillance signal. Admissions and occupancy are shown on different scales. (Bottom left) Reservoir reports daily death data. (Right) Reservoir reports daily pathogen concentration in wastewater as well as the number of outpatient consultations for the disease (syndromic).
    }
    \label{fig:surveillance}
\end{figure}

\subsection*{Example 2: Model structure, parameters, and initial conditions}

The following examples show how users can specify their own configurations. In particular, \texttt{model\_structure()} can be used to build an SIS model, where recovery confers no immunity and individuals return directly to the susceptible population:

\begin{lstlisting}[style=Rstyle]
library("Reservoir")

sis <- model_structure(
  has_latent  = FALSE,   # infection is immediately infectious
  has_asymptomatic = FALSE,   # a single infectious class
  immunity = "none",  # recovery returns to S
  has_isolation  = FALSE # infectious individuals are not isolated
)

\end{lstlisting}

Disease parameters are supplied through \texttt{reservoir\_config()}. Each field accepts a single value, which sets it for every simulation, or a pair, which is read as a central 90\% interval from which each simulation draws its own:

\begin{lstlisting}[style=Rstyle]
cfg <- reservoir_config(
  r0_base = 2.5,  
  infectious_days = c(4, 10)    # sampled, 90% of draws in this range
)
\end{lstlisting}

Initial conditions are set through \texttt{population\_spec()}, either as fractions or as explicit counts. The fractional form is convenient when the population size is itself being sampled:

\begin{lstlisting}[style=Rstyle]
pop <- population_spec(
  num_days = 365,
  pop_size = 1e5,
  seed_frac = 1e-4,     # fraction infectious on day 1
  novel = FALSE,        # part of the population is already immune
  immune_frac = 0.30
)
\end{lstlisting}

The explicit form states the starting occupancy of each compartment directly. The total population is inferred from the sum, and the compartments are distributed across strata in the same proportions as the population:

\begin{lstlisting}[style=Rstyle]
pop <- population_spec(
  num_days = 365,
  compartments = c(S = 99000, Is = 1000)
)
\end{lstlisting}

\section*{Impact}

\noindent Reservoir addresses two constraints that limit progress in computational epidemiology: the scarcity of training data for methods that require large datasets, and the absence of ground truth labels in epidemiological datasets for quantities that cannot be observed (e.g., true underlying cases, $R_t$, or scenario trajectories).

%  \begin{figure}
%     \centering
%      \includegraphics[width=\textwidth,  height=\textheight, keepaspectratio]{figure_rt_benchmark.pdf}
%     \caption{\textbf{} \marisa{add caption} \marisa{Also add the Rt section to the paper!} \marisa{also, if we're going to do Rt benchmarking, we should see how sensitive each method is to various real world effects---all probably perform pretty well with perfect data? (Or maybe not) But probably degrade differently with each of the real world effects.}
%     }
%     \label{fig:rt_benchmark}
% \end{figure}

Reservoir's most prominent application to date is Mantis \cite{mantis}, a foundation model for infectious disease forecasting trained entirely on Reservoir data, which is now being used by public health agencies to support outbreak awareness in multiple countries. Mantis demonstrates one of our main intended use-cases: models can learn the structure of transmission from large-scale simulation data, without disease-specific fitting. 
%The clearest evidence of the Reservoir's impact is as a training resource for Mantis, a foundation model for infectious disease forecasting trained entirely on Reservoir data that generalized effectively to real-world tasks, achieving high out-of-the-box accuracy across 16 diseases \cite{mantis}. This is the most significant validation of the dataset: a model that learned only from simulations transferred successfully to real outbreaks. 
Mantis' approach---training on large-scale synthetic data, then transferring to real tasks---has shown increasing success across domains \cite{sgnns2025, tabpfn, sgnntheory, timepfn}, and the scale and mechanistic diversity of Reservoir make it suitable for training models on tasks other than forecasting, including real-time $R_t$ estimation, anomaly detection, and cross-modal translation (e.g., inferring true case trajectories from wastewater signals).

As noted earlier, Reservoir can also enable benchmarking established methods against known answers in controlled, progressively challenging conditions. In addition to evaluating existing methods, the simulator's configurability supports experiments that real data cannot, such as counterfactual intervention scenarios or surveillance-system design studies (e.g., which modality or sampling cadence best supports a given inference task).

\section*{Conclusions}
 
\noindent Reservoir is a large-scale, ground-truth-labeled simulator and dataset for epidemic modeling, and it is already supporting research and public health practice. Because every simulation has complete latent-state labels, Reservoir can be used for evaluating methods against known ground truth and for counterfactual and surveillance-design studies that real outbreak data cannot support.
 
Our work has several limitations. The current simulators do not include a range of features, such as behavioral feedback loops (e.g., risk-driven behavior change) or healthcare-capacity effects (e.g. limited hospital beds). Additionally, all compartments are simulated via tau leaping, though for the environmental pathogen compartment an ODE representation might be more appropriate to capture the more continuous nature of environmental pathogen concentration (this may lead to less realistic, more `chunky' stochasticity due to the model treating `packets' of virions as the unit of interest, though in many cases this may help to capture the larger noise that likely comes from variation in environmental conditions, water flow, etc.). Simulations also use synthetic contact matrices and population structures rather than empirical demographic data for specific countries, which means there is not direct geographic calibration. However, since the simulator architecture is modular, new transmission mechanisms, surveillance modalities, and intervention types can be added without restructuring the codebase, and future work could incorporate more realistic environmental models, spatial dynamics, and empirical contact and demographic data for location-specific scenario generation. By releasing the simulator alongside its data, we intend Reservoir to be a resource that evolves with the field's needs through community contributions. 

\section*{Acknowledgments}
We want to thank Stephanie Iorga for testing the original versions of Reservoir and assisting with usability. This project was made possible by the Insight Net cooperative agreement with University of Michigan (5NU38FT000002-02-00) from the CDC’s Center for Forecasting and Outbreak Analytics (CDC-RFA-FT-23-0069). Its contents are solely the responsibility of the authors and do not necessarily represent the official
views of the Centers for Disease Control and Prevention. 

\section*{Author Contributions}
CD conceived of the work, co-led the study, developed the code, and initial draft of the manuscript. RM co-led the study, developed the code, developed figure visualizations, and initial draft of the manuscript. CD, MCE, and RM all contributed to the analysis plan and made substantial edits and revisions for the final manuscript.

\bibliographystyle{unsrt}
\bibliography{references}

@article{koopman2004modeling,
  title={Modeling infection transmission},
  author={Koopman, Jim},
  journal={Annu. Rev. Public Health},
  volume={25},
  number={1},
  pages={303--326},
  year={2004},
  publisher={Annual Reviews}
}

@article{alphafold,
  title={Highly accurate protein structure prediction with AlphaFold},
  author={Jumper, John and Evans, Richard and Pritzel, Alexander and Green, Tim and Figurnov, Michael and Ronneberger, Olaf and Tunyasuvunakool, Kathryn and Bates, Russ and Žídek, Augustin and Potapenko, Anna and others},
  journal={Nature},
  volume={596},
  number={7873},
  pages={583--589},
  year={2021},
  publisher={Nature Publishing Group},
  doi={10.1038/s41586-021-03819-2}
}

@article{pdb,
    author = {{wwPDB consortium}},
    title = "{Protein Data Bank: the single global archive for 3D macromolecular structure data}",
    journal = {Nucleic Acids Research},
    volume = {47},
    number = {D1},
    pages = {D520-D528},
    year = {2019},
    month = {01},
}

@article{barros2022causal,
  title={A causal inference approach for estimating effects of non-pharmaceutical interventions during Covid-19 pandemic},
  author={Barros, Vesna and Manes, Itay and Akinwande, Victor and Cintas, Celia and Bar-Shira, Osnat and Ozery-Flato, Michal and Shimoni, Yishai and Rosen-Zvi, Michal},
  journal={PLOS ONE},
  year={2022},
  month={09},
  publisher={Public Library of Science},
}

@article{ganyani2020estimating,
  title={Estimating the generation interval for coronavirus disease (COVID-19) based on symptom onset data, March 2020},
  author={Ganyani, Tapiwa and Kremer, C{\'e}cile and Chen, Dongxuan and Torneri, Andrea and Faes, Christel and Wallinga, Jacco and Hens, Niel},
  journal={Eurosurveillance},
  volume={25},
  number={17},
  pages={2000257},
  year={2020},
  month={04},
}

@article{mossong2008polymod,
  title={Social Contacts and Mixing Patterns Relevant to the Spread of Infectious Diseases},
  author={Mossong, Jo{\"e}l and others},
  journal={PLoS Medicine},
  publisher={Public Library of Science},
}

@article{cao2007adaptive,
  title={Adaptive explicit-implicit tau-leaping method with automatic tau selection},
  author={Cao, Yang and Gillespie, Daniel T and Petzold, Linda R},
  journal={The Journal of Chemical Physics},
  volume={126},
  number={22},
  pages={224101},
  year={2007},
}

@article{gillespie1977exact,
  title={Exact stochastic simulation of coupled chemical reactions},
  author={Gillespie, Daniel T},
  journal={The Journal of Physical Chemistry},
  volume={81},
  number={25},
  pages={2340--2361},
  year={1977},
  publisher={American Chemical Society},
}

@article{tabpfn,
  title={Accurate predictions on small data with a tabular foundation model},
  author={Hollmann, Noah and M{\"u}ller, Samuel and others},
  journal={Nature},
  volume={637},
  pages={319--326},
  year={2025},
}

@inproceedings{timepfn,
    title={Time{PFN}: Effective Multivariate Time Series Forecasting with Synthetic Data},
    author={Ege Onur Taga and M. Emrullah Ildiz and Samet Oymak},
    booktitle={Proceedings of the AAAI Conference on Artificial Intelligence},
    year={2025},
}

@article{gpt2,
  title={Language Models are Unsupervised Multitask Learners},
  author={Radford, Alec and Wu, Jeffrey and Child, Rewon and Luan, David and Amodei, Dario and Sutskever, Ilya},
  journal={OpenAI Technical Report},
  year={2019},
}

@article{gpt3,
  title={Language Models are Few-Shot Learners},
  author={Brown, Tom B. and Mann, Benjamin and Ryder, Nick and Subbiah, Melanie and Kaplan, Jared D. and Dhariwal, Prafulla and Neelakantan, Arvind and Shyam, Pranav and Sastry, Girish and Askell, Amanda and others},
  journal={Advances in Neural Information Processing Systems},
  volume={33},
  pages={1877--1901},
  year={2020},
}

@article{scalinglaws,
  title={Scaling Laws for Neural Language Models},
  author={Kaplan, Jared and McCandlish, Sam and Henighan, Tom and Brown, Tom B. and Chess, Benjamin and Child, Rewon and Gray, Scott and Radford, Alec and Wu, Jeffrey and Amodei, Dario},
  journal={arXiv preprint arXiv:2001.08361},
  year={2020},
  url={https://arxiv.org/abs/2001.08361}
}

@article{pew2024publichealth,
  title={State Public Health Data Reporting Policies and Practices Vary Widely},
  author={{The Pew Charitable Trusts}},
  journal={Advance Health \& Well-Being},
  year={2024},
}

@article{cao2006efficient,
  author  = {Cao, Yang and Gillespie, Daniel T. and Petzold, Linda R.},
  title   = {Efficient step size selection for the tau-leaping simulation method},
  journal = {The Journal of Chemical Physics},
  volume  = {124},
  year    = {2006},
}

@article{sgnns2025,
  title     = {Training neural networks on mechanistic simulations improves scientific inference},
  author    = {Dudley, Carson and Magdaleno, Reiden and Harding, Christopher and Eisenberg, Marisa},
  journal   = {Scientific Reports},
  year      = {2026},
  publisher = {Nature Publishing Group},
}

@misc{delphi_epidata,
  author       = {Farrow, David C. and Brooks, Logan C. and Rumack, Aaron and Tibshirani, Ryan J. and Rosenfeld, Roni},
  title        = {Delphi Epidata API},
  year         = {2015},
  howpublished = {\url{https://github.com/cmu-delphi/delphi-epidata}},
  note         = {Carnegie Mellon University, Delphi Research Group}
}

@article{project_tycho,
  title        = {Project Tycho 2.0: a repository to improve the integration and reuse of data for global population health},
  author       = {van Panhuis, Willem G. and Cross, Anne and Burke, Donald S. and others},
  journal      = {Journal of the American Medical Informatics Association},
  volume       = {25},
  number       = {12},
  pages        = {1608--1617},
  year         = {2018},
}

@article{mantis,
    title={Mantis: A Simulation-Grounded Foundation Model for Disease Forecasting}, 
    author={Carson Dudley and Reiden Magdaleno and Christopher Harding and Ananya Sharma and Marisa Eisenberg},
    year={2025},
    journal={arXiv preprint arXiv:2508.12260},
    url={https://arxiv.org/abs/2508.12260},
}

@article{gao2025early,
  title={Early detection of disease outbreaks and non-outbreaks using incidence data: A framework using feature-based time series classification and machine learning},
  author={Gao, Shan and Chakraborty, Amit K and Greiner, Russell and Lewis, Mark A and Wang, Hao},
  journal={PLoS Computational Biology},
  volume={21},
  number={2},
  year={2025},
  month={Feb},
}

@misc{mu2025counterfactualprobabilisticdiffusionexpert,
      title={Counterfactual Probabilistic Diffusion with Expert Models}, 
      author={Wenhao Mu and Zhi Cao and Mehmed Uludag and Alexander Rodríguez},
      year={2025},
      eprint={2508.13355},
      archivePrefix={arXiv},
      primaryClass={cs.LG},
      url={https://arxiv.org/abs/2508.13355}, 
}

@misc{deepgleam,
      title={DeepGLEAM: A hybrid mechanistic and deep learning model for COVID-19 forecasting}, 
      author={Dongxia Wu and Liyao Gao and Xinyue Xiong and Matteo Chinazzi and Alessandro Vespignani and Yi-An Ma and Rose Yu},
      year={2021},
      eprint={2102.06684},
      archivePrefix={arXiv},
      primaryClass={cs.LG},
      url={https://arxiv.org/abs/2102.06684}, 
}

@article{sgnntheory,
  title={Learning From Simulators: A Theory of Simulation-Grounded Learning},
  author={Carson Dudley and Marisa Eisenberg},
  journal={arXiv preprint arXiv:2509.18990},
  year={2025},
  url={https://arxiv.org/abs/2509.18990}
}

@article{hakki2022onset,
  title={Onset and window of SARS-CoV-2 infectiousness and temporal correlation with symptom onset: a prospective, longitudinal, community cohort study},
  author={Hakki, Seran and Zhou, Jie and Jonnerby, Jakob and Singanayagam, Anika and Barnett, Jack L and Madon, Kieran J and Koycheva, Aleksandra and Kelly, Christine and Houston, Hamish and Nevin, Sean and others},
  journal={The Lancet Respiratory Medicine},
  volume={10},
  number={11},
  pages={1061--1073},
  year={2022},
  publisher={Elsevier}
}

@article{gani2004epidemiologic,
  title={Epidemiologic determinants for modeling pneumonic plague outbreaks},
  author={Gani, Raymond and Leach, Steve},
  journal={Emerging infectious diseases},
  volume={10},
  number={4},
  pages={608},
  year={2004}
}

@article{ke2021estimating,
  title={Estimating the reproductive number R0 of SARS-CoV-2 in the United States and eight European countries and implications for vaccination},
  author={Ke, Ruian and Romero-Severson, Ethan and Sanche, Steven and Hengartner, Nick},
  journal={Journal of theoretical biology},
  volume={517},
  pages={110621},
  year={2021},
  publisher={Elsevier}
}

@article{nikbakht2019comparison,
  title={Comparison of methods to estimate basic reproduction number (R0) of influenza, using Canada 2009 and 2017-18 A (H1N1) data},
  author={Nikbakht, Roya and Baneshi, Mohammad Reza and Bahrampour, Abbas and Hosseinnataj, Abolfazl},
  journal={Journal of research in medical sciences: the official journal of Isfahan University of Medical Sciences},
  volume={24},
  pages={67},
  year={2019}
}

@article{kiang2025modeling,
  title={Modeling reemergence of vaccine-eliminated infectious diseases under declining vaccination in the US},
  author={Kiang, Mathew V and Bubar, Kate M and Maldonado, Yvonne and Hotez, Peter J and Lo, Nathan C},
  journal={JAMA},
  volume={333},
  number={24},
  pages={2176--2187},
  year={2025}
}

@article{xin2022estimating,
  title={Estimating the latent period of coronavirus disease 2019 (COVID-19)},
  author={Xin, Hualei and Li, Yu and Wu, Peng and Li, Zhili and Lau, Eric HY and Qin, Ying and Wang, Liping and Cowling, Benjamin J and Tsang, Tim K and Li, Zhongjie},
  journal={Clinical Infectious Diseases},
  volume={74},
  number={9},
  pages={1678--1681},
  year={2022},
  publisher={Oxford University Press US}
}

@article{edridge2020seasonal,
  title={Seasonal coronavirus protective immunity is short-lasting},
  author={Edridge, Arthur WD and Kaczorowska, Joanna and Hoste, Alexis CR and Bakker, Margreet and Klein, Michelle and Loens, Katherine and Jebbink, Maarten F and Matser, Amy and Kinsella, Cormac M and Rueda, Paloma and others},
  journal={Nature medicine},
  volume={26},
  number={11},
  pages={1691--1693},
  year={2020},
  publisher={Nature Publishing Group US New York}
}

@article{amanna2007duration,
  title={Duration of humoral immunity to common viral and vaccine antigens},
  author={Amanna, Ian J and Carlson, Nichole E and Slifka, Mark K},
  journal={New England Journal of Medicine},
  volume={357},
  number={19},
  pages={1903--1915},
  year={2007},
  publisher={Mass Medical Soc}
}

@article{sah2021asymptomatic,
  title={Asymptomatic SARS-CoV-2 infection: A systematic review and meta-analysis},
  author={Sah, Pratha and Fitzpatrick, Meagan C and Zimmer, Charlotte F and Abdollahi, Elaheh and Juden-Kelly, Lyndon and Moghadas, Seyed M and Singer, Burton H and Galvani, Alison P},
  journal={Proceedings of the National Academy of Sciences},
  volume={118},
  number={34},
  pages={e2109229118},
  year={2021},
  publisher={National Academy of Sciences}
}

@article{leung2015fraction,
  title={The fraction of influenza virus infections that are asymptomatic: a systematic review and meta-analysis},
  author={Leung, Nancy HL and Xu, Cuiling and Ip, Dennis KM and Cowling, Benjamin J},
  journal={Epidemiology (Cambridge, Mass.)},
  volume={26},
  number={6},
  pages={862},
  year={2015}
}

@article{buitrago2022occurrence,
  title={Occurrence and transmission potential of asymptomatic and presymptomatic SARS-CoV-2 infections: Update of a living systematic review and meta-analysis},
  author={Buitrago-Garcia, Diana and Ipekci, Aziz Mert and Heron, Leonie and Imeri, Hira and Araujo-Chaveron, Lucia and Arevalo-Rodriguez, Ingrid and Ciapponi, Agust{\'\i}n and Cevik, Muge and Hauser, Anthony and Alam, Muhammad Irfanul and others},
  journal={PLoS medicine},
  volume={19},
  number={5},
  pages={e1003987},
  year={2022},
  publisher={Public Library of Science San Francisco, CA USA}
}

@article{wearing2005appropriate,
  title={Appropriate models for the management of infectious diseases},
  author={Wearing, Helen J and Rohani, Pejman and Keeling, Matt J},
  journal={PLoS medicine},
  volume={2},
  number={7},
  pages={e174},
  year={2005},
  publisher={Public Library of Science San Francisco, USA}
}

@article{liu2020reviewing,
  title={Reviewing estimates of the basic reproduction number for dengue, Zika and chikungunya across global climate zones},
  author={Liu, Ying and Lillepold, Kate and Semenza, Jan C and Tozan, Yesim and Quam, Mikkel BM and Rockl{\"o}v, Joacim},
  journal={Environmental Research},
  volume={182},
  pages={109114},
  year={2020},
  publisher={Elsevier}
}

@article{mukandavire2011estimating,
  title={Estimating the reproductive numbers for the 2008--2009 cholera outbreaks in Zimbabwe},
  author={Mukandavire, Zindoga and Liao, Shu and Wang, Jin and Gaff, Holly and Smith, David L and Morris Jr, J Glenn},
  journal={Proceedings of the National Academy of Sciences},
  volume={108},
  number={21},
  pages={8767--8772},
  year={2011},
  publisher={National Academy of Sciences}
}

@article{kucharski2020effectiveness,
  title={Effectiveness of isolation, testing, contact tracing, and physical distancing on reducing transmission of SARS-CoV-2 in different settings: a mathematical modelling study},
  author={Kucharski, Adam J and Klepac, Petra and Conlan, Andrew JK and Kissler, Stephen M and Tang, Maria L and Fry, Hannah and Gog, Julia R and Edmunds, W John and Emery, Jon C and Medley, Graham and others},
  journal={The Lancet infectious diseases},
  volume={20},
  number={10},
  pages={1151--1160},
  year={2020},
  publisher={Elsevier}
}

@article{ambrosioni2021primary,
  title={Primary HIV-1 infection in users of pre-exposure prophylaxis},
  author={Ambrosioni, Juan and Petit, Elisa and Liegeon, Geoffroy and Laguno, Montserrat and Mir{\'o}, Jos{\'e} M},
  journal={The Lancet HIV},
  volume={8},
  number={3},
  pages={e166--e174},
  year={2021},
  publisher={Elsevier}
}

@article{arciuolo2017effectiveness,
  title={Effectiveness of measles vaccination and immune globulin post-exposure prophylaxis in an outbreak setting—New York City, 2013},
  author={Arciuolo, Robert J and Jablonski, Rachel R and Zucker, Jane R and Rosen, Jennifer B},
  journal={Clinical Infectious Diseases},
  volume={65},
  number={11},
  pages={1843--1847},
  year={2017},
  publisher={Oxford University Press US}
}

@article{auranen2023efficacy,
  title={Efficacy and effectiveness of case isolation and quarantine during a growing phase of the COVID-19 epidemic in Finland},
  author={Auranen, Kari and Shubin, Mikhail and Erra, Elina and Isosomppi, Sanna and Kontto, Jukka and Leino, Tuija and Lukkarinen, Timo},
  journal={Scientific reports},
  volume={13},
  number={1},
  pages={298},
  year={2023},
  publisher={Nature Publishing Group UK London}
}

\begin{appendices}

\section{Modeling Framework and Numerical Method}
\label{sec:framework}

The simulator implements a structured stochastic compartmental model for infectious diseases transmitted by three mechanisms: direct human-to-human contact (Section \ref{sec:h2h}), a vector intermediary (Section \ref{sec:vector}), and an environmental pathogen reservoir (Section \ref{sec:env}). All three share an identical human compartment structure, an identical reaction-based formulation, and an identical numerical integrator. They differ only in the additional states they have (e.g., vector states) and how those affect the force of infection calculation. This section describes the overall simulator structure and numerical methods used.

 \begin{figure}[H]
    \centering
     \includegraphics[width=\textwidth,  height=\textheight, keepaspectratio]{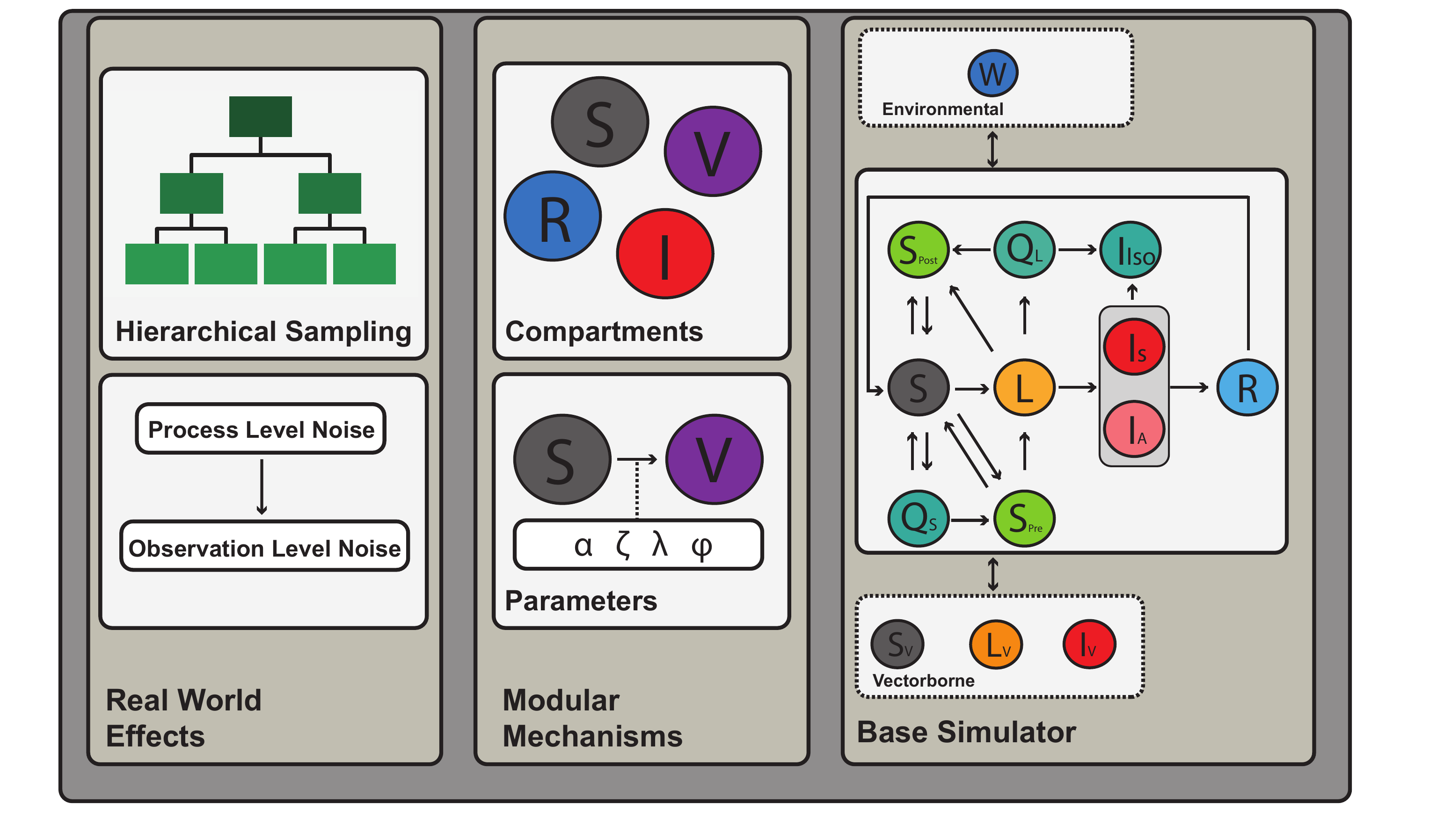}
 \caption{\textbf{Dataset Generation Workflow.} 
    }
    \label{fig:conceptual}
\end{figure}

\subsection*{Continuous-time Markov chain}

The population is stratified by $P$ subpopulations and $A$ age groups, giving $P \times A$ demographic strata. The epidemic state is a vector $\mathbf{x}(t) \in \mathbb{Z}_{\ge 0}^{M}$ of non-negative integer counts. Time is continuous and the system evolves as a continuous-time Markov chain (CTMC): a finite set of reaction channels $\{R_k\}_{k=1}^{K}$ each fire at random with state-dependent propensity $a_k(\mathbf{x},t)$ (events per day), and firing $R_k$ updates the state by a vector $\boldsymbol{\nu}_k$,
\begin{equation}
  \mathbf{x} \;\longrightarrow\; \mathbf{x} + \boldsymbol{\nu}_k .
\end{equation}

All transitions follow the law of mass action with the time-varying rate constants described in Section \ref{sec:forcing}. Because the exact propensities depend on the force of infection, which is a nonlinear function of the full state, they are recomputed at every numerical step.

\begin{table}[H]
\centering
\begin{tabular}{cll}
\toprule
Variable & Description \\
\midrule
$S$        & Susceptible \\
$S_\text{prep}$ & Susceptible, on PrEP \\
$S_\text{post}$ & Post-PEP protected (temporarily immune) \\
$Q_S$      & Susceptible in quarantine \\
$L$        & Latent (exposed, not yet infectious) \\
$Q_L$      & Latent in quarantine \\
$I_s$      & Infectious, symptomatic \\
$I_a$      & Infectious, asymptomatic \\
$I_{iso}$   & Infectious, isolated \\
$R$        & Recovered (waning stage 1, always present) \\
$R_2$      & Recovered (waning stage 2, used if $n_w \ge 2$) \\
$R_3$      & Recovered (waning stage 3, used if $n_w \ge 3$) \\
$C$        & Cumulative infections (tracker, not a true compartment) \\
\bottomrule
\end{tabular}
\caption{The 13 human variables that can be tracked.}
\label{tab:blocks}
\end{table}

The total living population of a stratum $(p,a)$ is:
\begin{equation}
  N_{p,a} = S_{p,a} + S_{\text{prep},p,a} + S_{\text{post},p,a}
          + Q_{S,p,a} + L_{p,a} + Q_{L,p,a}
          + I_{s,p,a} + I_{a,p,a} + I_{iso,p,a}
          + R_{p,a} + R_{2,p,a} + R_{3,p,a}.
\end{equation}

\subsection*{Time-varying forcing and demographic drivers}
\label{sec:forcing}

Each day, the simulator calculates an effective transmission rate $\beta_\text{eff}(t)$, which takes the transmission rate, $\beta$, for the current wave and scales it by a seasonal multiplier:
\begin{equation}
  \beta_\text{eff}(t) \;=\; \beta_\text{wave}(t)\, s(t).
\end{equation}

\paragraph{Epidemic waves.} Each simulation can be partitioned into multiple waves. Our default configuration chooses the number of waves at random between 0 and 4. After the number of waves is selected, we select when the wave will change ($\texttt{wave\_change\_days}$) and the new transmissibility parameters ($\texttt{wave\_change\_betas}$) independently. The first wave uses the baseline $\beta$. At each change day $d \in \texttt{wave\_change\_days}$, $\beta_\text{wave}(t)$ switches to the corresponding value in \texttt{wave\_change\_betas}, so $\beta_\text{wave}(t)$ is a right-continuous step function of the day index.

\paragraph{Seasonality.}
When seasonal forcing is enabled, the multiplier is a sum of up to three cosine
harmonics with optional annual peak jitter and multiplicative daily noise,
\begin{equation}
  s(t) = \max\!\Bigl(0,\; 1 + \sum_{h=1}^{n_h} A_h
            \cos\!\Bigl(\tfrac{2\pi\,(t - \phi_h - \xi_{\text{yr}(t)})}{T_h}\Bigr)\Bigr)\,
          \,\zeta_t,
  \label{eq:seasonal}
\end{equation}
where $A_h$, $\phi_h$ and $T_h$ are the amplitude, phase offset (days) and period (days) of harmonic $h$, $\xi_{\text{yr}(t)} \sim \mathcal{N}(0, \sigma_\text{jit}^2)$ is a per-calendar-year peak shift, and $\zeta_t$ is daily noise clipped to $[0.5,2]$ drawn from $\mathcal{N}(1, \sigma_\text{day}^2)$. With no harmonics but nonzero daily noise, $s(t)=\zeta_t$; otherwise $s(t)\equiv 1$.

\paragraph{Super-spreading.}
For dataset generation an additional multiplicative factor is applied to $\beta_\text{eff}(t)$ on a random subset of days to emulate super-spreading. On super-spreading event days $\beta_\text{eff}$ is scaled by a draw from $\text{Gamma}(\kappa_\text{ss},\theta_\text{ss})$. We want super-spreading events to be present to make the simualtions more realistic, but we do not want these multiplications to increase the realized transmissibility of the disease overall, so after the super-spreading event day multipliers are selected, the whole multiplier series is rescaled by its expectation so that $\mathbb{E}[\text{multiplier}]=1$ and the mean transmission rate is preserved.

\paragraph{Interventions.} Since interventions may sometimes be used, but not always, the presence of each intervention is a time varying parameter: for isolation $\nu_\text{iso}(t)$, quarantine $\nu_\text{quar}(t)$, PEP $\nu_\text{pep}(t)$ and PrEP efficacy $\varepsilon_\text{prep}(t)$.

\paragraph{Importation.}
At the start of each day, before we simulate any transmission, external infections are injected from the susceptible population to simulate case importation:
\begin{equation}
  n_\text{imp}(t) \sim \text{Poisson}\bigl(\lambda_\text{imp}\bigr).
\end{equation}
Importation often keeps endemic runs from going extinct between waves. 

\paragraph{Demography.}
Births enter $S$ at per-capita daily rate $b$, and every living compartment is subject to background mortality at per-capita daily rate $d$.

\subsection*{Structural model variants}
\label{sec:structure}

The compartment set in Table \ref{tab:blocks} is the maximal one. Individual simulations usually use a strict subset of it: before parameters are drawn, a structure is sampled that determines which compartments are included and which reaction channels are built. Structural flags change the reaction network itself rather than setting a rate to zero, so two runs in the same dataset may be governed by different compartmental models. The realized structure is recorded with every run.

\paragraph{Latent stage.} When the latent stage is present (the default), infection goes into $L$ and progression out of $L$ branches into $I_s$ and $I_a$ with probabilities $1-p_a$ and $p_a$. When it is absent, $L$ and $Q_L$ are unreachable and infection branches directly at the point of transmission,

\begin{align}
  S_{p,a} \to I_{s,p,a}, \quad C_{p,a}+1
    &\qquad \text{rate: } \lambda_{p,a}\,(1-p_a)\, S_{p,a}, \\  S_{p,a} \to I_{a,p,a}, \quad C_{p,a}+1 &\qquad \text{rate: } \lambda_{p,a}\, p_a\, S_{p,a},
\end{align}

with the analogous pair out of $S_\text{prep}$ carrying the additional factor $(1-\varepsilon_\text{prep}(t))$. The total infection flux is unchanged. Reactions involving $L$ or $Q_L$ are removed ($L \to Q_L$, $L \to S_\text{post}$, $Q_L \to S_\text{post}$, $L \to I_s$, $L \to I_a$, $Q_L \to I_a$, $Q_L \to I_{iso}$) and two infection channels are added. Because PEP acts only on $L$ and $Q_L$, it is not used with this structure and $S_\text{post}$ is always empty. Imported cases enter $I_s$ or $I_a$, split by $p_a$, rather than $L$.

\paragraph{Immunity.} Three regimes are available. Under waning immunity (the default) $\omega > 0$ and recovered individuals return to $S$ through $n_w$ stages as in Section \ref{sec:h2h}. Under permanent immunity $\omega = 0$ and the $R$ compartment is permanent. Under no immunity the model is SIS: every recovery transition goes straight to $S$, and preexisting immunity is forced to zero regardless of the initial-condition specification (Section \ref{sec:initcond}).

\paragraph{Recovery.} When recovery is disabled, $\gamma = 0$ and the model is SI (or SLI with a latent stage). The infectious period is unbounded, $R_0$ diverges, and $\beta$ must then be supplied directly rather than derived. $R_0$ is recorded as undefined for such runs.

\paragraph{Other flags.} The asymptomatic class is disabled parametrically by setting $p_a = \alpha = \kappa = 0$, which leaves the $I_a$ channels in the network at zero propensity. Isolation, quarantine, PEP, vaccination/PrEP, demography and importation are also switched on or off per run. When disabled the corresponding rate constants are zero.

\paragraph{Realized networks.} With a latent stage and three waning stages the engine builds 37 reactions per stratum. With a single waning stage the count is 33. Removing the latent stage reduces each of these by five.

\paragraph{Mixtures.} Each structural flag accepts a probability rather than a Boolean, in which case it is drawn independently per simulation. A single dataset can therefore span SIR, SEIR, SEIAR, SIS, SIRS and SI model structures. Every run carries its realiszd flags and a label (for example \code{SLIARS}) alongside its parameters, so structure can be conditioned on, stratified by, or used as a supervision target.

\subsection*{Interventions}
\label{sec:interv}

Interventions enter the model in two ways. Measures that move individuals between compartments (isolation, quarantine, PEP, and PrEP/vaccination) act through the daily rates $\nu_\text{iso}(t)$, $\nu_\text{quar}(t)$, $\nu_\text{pep}(t)$, $\rho_\text{start}(t)$ and $\varepsilon_\text{prep}(t)$. Measures that change contact structure (i.e., school and workplace closure) act multiplicatively on the transmission rate and do not move anyone between compartments. Whether a given simulation receives any intervention at all is itself sampled.

\paragraph{Scheduled mode.} All active measures share a single window $[t_\text{on}, t_\text{off}]$, with $t_\text{on}$ drawn as a fraction of the horizon and the duration drawn independently. Each driver takes its baseline value outside the window and its active value inside it. Contact reduction is applied as a constant multiplier $(1 - c_\text{school})(1 - c_\text{work})$ on $\beta_\text{eff}(t)$.

\paragraph{Reactive mode.} Interventions are triggered by the epidemic itself rather than by the calendar. The controller carries a binary state $\iota_t \in \{0,1\}$ and a countdown, updated once per day before any transmission is simulated. Writing $y_{t-1}$ for the previous day's new infections, $\theta_\text{on}$ and $\theta_\text{off}$ for the on and off thresholds, and $\delta$ for the detection lag:

\begin{itemize}
  \item If $\iota = 0$ and no intervention is pending, intervention begins the first day $y_{t-1} \geq \theta_\text{on}$, with the countdown set to $\delta$.
  \item A countdown counts down each day and the intervention ends on reaching zero. It is not disarmed if $y_{t-1}$ falls below $\theta_\text{on}$, so a single threshold crossing commits the system to an intervention $\delta$ days later.
  \item If $\iota = 1$, the intervention is released the first day $y_{t-1} \leq \theta_\text{off}$. Because the intervention is re-evaluated daily a single run may cycle through several on/offs.
\end{itemize}

\subsection*{Adaptive tau-leaping integrator}
\label{sec:tau}

The state is advanced one day at a time. Within a day the elapsed time $t_\text{in} \in [0,1)$ is integrated with an explicit adaptive tau-leaping scheme using the \cite{cao2006efficient, cao2007adaptive}, which separates critical reactions (those at risk of driving a low-count compartment negative) from non-critical reactions, and falls back on the exact stochastic simulation algorithm \cite{gillespie1977exact}. At each within-day step the force of infection and all propensities are recomputed from the current state before the step size is chosen.

\paragraph{Criticality.} A reaction $R_k$ is critical at state $\mathbf{x}$ if it consumes any compartment whose current count is below 10. Critical reactions are excluded from the Poisson leap and handled one event at a time. Because the model is stratified among many populations, low-count compartments are common at the beginning and end of an outbreak and in small populations, so a substantial fraction of the reaction set is critical. At most one critical reaction fires per step.

\paragraph{Termination.} The within-day loop exits when the time advances past the end of the day, when the total propensity vanishes, or when a step limit of $10^7$ within-day iterations is reached, in which case a warning is given. In the last two cases the remainder of the day is not integrated.

%==============================================================================
\section{Human-to-Human Simulator}
\label{sec:h2h}
%==============================================================================

\subsection*{Compartments}

As a reminder, the state vector contains $P \times A \times 13$ elements, organised into 13 blocks of size $P \times A$. Each block corresponds to one epidemiological compartment. Let $X_{p,a}$ denote the count in compartment $X$ for subpopulation $p \in \{1,\ldots,P\}$ and age group $a \in \{1,\ldots,A\}$ (compartments are defined in table \ref{tab:blocks}).

The total population in stratum $(p, a)$ is:
\begin{equation}
  N_{p,a} = S_{p,a} + S_{\text{prep},p,a} + S_{\text{post},p,a}
          + Q_{S,p,a} + L_{p,a} + Q_{L,p,a}
          + I_{s,p,a} + I_{a,p,a} + I_{so,p,a}
          + R_{p,a} + R_{2,p,a} + R_{3,p,a}
\end{equation}

\subsection*{Force of Infection}

Transmission is \textbf{frequency dependent}. The force of infection (FOI) for stratum $(p, a)$ is:

\begin{equation}
  \lambda_{p,a}(t)
  = \sum_{q=1}^{P} M_{pq}
    \sum_{b=1}^{A} \beta_\text{eff}(t)\, f_\text{interv}(t) \cdot C_{ab}
    \cdot \frac{I_{s,q,b} + \alpha\, I_{a,q,b}}{N_{q,b}}
\end{equation}

where:
\begin{itemize}
  \item $M_{pq}$ is the $(p,q)$ entry of the $P \times P$ inter-population
        mixing matrix,
  \item $C_{ab}$ is the $(a,b)$ entry of the $A \times A$ age-structured
        contact matrix,
  \item $\beta_\text{eff}(t)$ is the time-varying effective transmission rate
        (daily driver vector supplied at runtime),
  \item $\alpha \in [0,1]$ is the relative transmissibility of asymptomatic individuals (\texttt{asymp\_trans}),
  \item $I_{iso}$ does \emph{not} contribute to transmission (isolated individuals have no contacts), so $I_{iso}$ is not included in the numerator, but it does contribute to the denominator.
    \item $f_\text{interv}(t)$ is the scheduled contact-reduction multiplier:  $(1-c_\text{school})(1-c_\text{work})$ inside the intervention window and $1$ outside it (Section \ref{sec:interv})
\end{itemize}

\subsection*{Transition Rates}

All transitions are listed below. Each rate is a propensity (events per unit time) for stratum $(p, a)$. Time-varying quantities (e.g., $\beta_\text{eff}(t)$ and $\nu_\text{iso}(t)$) are pre-computed for each day and supplied to the simulator as vectors. Under the SIS structure (\texttt{sis\_mode}; Section \ref{sec:structure}), all transitions described below as ending in $R_{p,a}$ instead return individuals to $S_{p,a}$. The $R$ compartment is still coded in the state vector for consistency, but remains empty.

\subsubsection*{Infection}

\begin{align}
  S_{p,a} \to L_{p,a}, \quad C_{p,a} + 1
    &\qquad \text{rate: } \lambda_{p,a}\, S_{p,a} \\[4pt]
  S_{\text{prep},p,a} \to L_{p,a}, \quad C_{p,a} + 1
    &\qquad \text{rate: } \lambda_{p,a}\,(1 - \varepsilon_\text{prep})\,
      S_{\text{prep},p,a}
\end{align}

where $\varepsilon_\text{prep}(t) \in [0,1]$ is the daily PrEP efficacy.

\subsubsection*{PrEP Dynamics}

\begin{align}
  S_{p,a} \to S_{\text{prep},p,a}
    &\qquad \text{rate: } \rho_\text{start}(t)\, S_{p,a} \\[4pt]
  S_{\text{prep},p,a} \to S_{p,a}
    &\qquad \text{rate: } \rho_\text{stop}(t)\, S_{\text{prep},p,a}
\end{align}

\subsubsection*{Susceptible Quarantine}

\begin{align}
  S_{p,a} \to Q_{S,p,a}
    &\qquad \text{rate: } \nu_\text{quar}(t)\, S_{p,a} \\[4pt]
  Q_{S,p,a} \to S_{p,a}
    &\qquad \text{rate: } \nu_\text{leave}(t)\, Q_{S,p,a}
      \quad \text{(no PrEP on exit)} \\[4pt]
  Q_{S,p,a} \to S_{\text{prep},p,a}
    &\qquad \text{rate: } \nu_\text{leave,prep}(t)\, Q_{S,p,a}
      \quad \text{(resume PrEP on exit)}
\end{align}

\subsubsection*{Latent Quarantine}

\begin{equation}
  L_{p,a} \to Q_{L,p,a}
    \qquad \text{rate: } \nu_\text{quar}(t)\, L_{p,a}
\end{equation}

\subsubsection*{PEP (Post-Exposure Prophylaxis)}

PEP can clear an active latent infection, transitioning the individual to a temporarily protected post-PEP state:

\begin{align}
  L_{p,a} \to S_{\text{post},p,a}
    &\qquad \text{rate: } \nu_\text{pep}(t)\, L_{p,a} \\[4pt]
  Q_{L,p,a} \to S_{\text{post},p,a}
    &\qquad \text{rate: } \nu_\text{pep}(t)\, Q_{L,p,a} \\[4pt]
  S_{\text{post},p,a} \to S_{p,a}
    &\qquad \text{rate: } \omega_\text{post}\, S_{\text{post},p,a}
\end{align}

\subsubsection*{Disease Progression from $L$}

\begin{align}
  L_{p,a} \to I_{s,p,a}
    &\qquad \text{rate: } \sigma\,(1 - p_a)\, L_{p,a} \\[4pt]
  L_{p,a} \to I_{a,p,a}
    &\qquad \text{rate: } \sigma\, p_a\, L_{p,a}
\end{align}

where $\sigma$ is the progression rate out of the latent compartment ($= 1/\text{mean latent period}$) and $p_a$ is the probability of remaining asymptomatic.

\subsubsection*{Disease Progression from $Q_L$}

Quarantined latents progress directly to isolation or to asymptomatic:

\begin{align}
  Q_{L,p,a} \to I_{a,p,a}
    &\qquad \text{rate: } \sigma\, p_a\, Q_{L,p,a} \\[4pt]
  Q_{L,p,a} \to I_{so,p,a}
    &\qquad \text{rate: } \sigma\,(1-p_a)\, Q_{L,p,a}
\end{align}

\subsubsection*{Asymptomatic Transitions}

\begin{align}
  I_{a,p,a} \to I_{s,p,a}
    &\qquad \text{rate: } \kappa\, I_{a,p,a} \\[4pt]
  I_{a,p,a} \to I_{so,p,a}
    &\qquad \text{rate: } \nu_\text{iso}(t)\, I_{a,p,a} \\[4pt]
  I_{a,p,a} \to R_{p,a}
    &\qquad \text{rate: } \gamma\, I_{a,p,a}
\end{align}

where $\kappa$ is the rate at which asymptomatics develop symptoms and $\nu_\text{iso}(t)$ is the daily isolation rate.

\subsubsection*{Symptomatic Transitions}

\begin{align}
  I_{s,p,a} \to I_{so,p,a}
    &\qquad \text{rate: } \nu_\text{iso}(t)\, I_{s,p,a} \\[4pt]
  I_{s,p,a} \to R_{p,a}
    &\qquad \text{rate: } \gamma\, I_{s,p,a}
\end{align}

\subsubsection*{Isolated Transitions}

\begin{equation}
  I_{so,p,a} \to R_{p,a}
    \qquad \text{rate: } \gamma\, I_{so,p,a}
\end{equation}

\subsubsection*{Waning Immunity}

Waning immunity is modeled with up to 3 staged compartments through a gamma distribution.

\paragraph{Single stage ($n_w = 1$, exponential waning):}
\begin{equation}
  R_{p,a} \to S_{p,a}
    \qquad \text{rate: } \omega\, R_{p,a}
\end{equation}

\paragraph{Two stages ($n_w = 2$):}
\begin{align}
  R_{p,a}   \to R_{2,p,a} &\qquad \text{rate: } \omega\, R_{p,a} \\
  R_{2,p,a} \to S_{p,a}   &\qquad \text{rate: } \omega\, R_{2,p,a}
\end{align}

\paragraph{Three stages ($n_w = 3$):}
\begin{align}
  R_{p,a}   \to R_{2,p,a} &\qquad \text{rate: } \omega\, R_{p,a} \\
  R_{2,p,a} \to R_{3,p,a} &\qquad \text{rate: } \omega\, R_{2,p,a} \\
  R_{3,p,a} \to S_{p,a}   &\qquad \text{rate: } \omega\, R_{3,p,a}
\end{align}

\subsubsection*{Demographics}

\begin{align}
  Births \to S_{p,a}
    &\qquad \text{rate: } b\, N_{p,a}
      \qquad \text{(births enter susceptible class)} \\[4pt]
  X_{p,a} \to Deaths
    &\qquad \text{rate: } d\, X_{p,a}
      \qquad \text{for all living compartments } X
\end{align}

where $b$ and $d$ are the daily per-capita birth and death rates respectively.

\subsubsection*{Importation}
External cases are injected daily before the tau-leaping loop:
\begin{equation}
  n_\text{imp}(t) \sim \text{Poisson}(\lambda_\text{imp})
\end{equation}
Imported individuals are drawn proportionally from susceptible cells and moved $S_{p,a} \to L_{p,a}$, incrementing $C_{p,a}$.

\subsection*{Parameters}

\begin{table}[H]
\centering
\begin{tabular}{lll}
\toprule
Parameter & Symbol & Description \\
\midrule
\multicolumn{3}{l}{\emph{Rate constants}}\\
\texttt{sigma}             & $\sigma$              & Latent progression rate ($1/\text{latent period}$) \\
\texttt{gamma}             & $\gamma$              & Recovery rate ($1/\text{infectious period}$); $0$ for SI models \\
\texttt{omega}             & $\omega$              & Waning rate per stage; mean immunity duration is $n_w/\omega$ \\
\texttt{p\_asym}           & $p_a$                 & Probability of asymptomatic infection \\
\texttt{asymp\_trans}      & $\alpha$              & Relative transmissibility of asymptomatics \\
\texttt{kappa}             & $\kappa$              & Rate $I_a \to I_s$ \\
\texttt{birth\_rate}       & $b$                   & Per-capita daily birth rate \\
\texttt{death\_rate}       & $d$                   & Per-capita daily death rate \\
\texttt{spost\_waning\_rate} & $\omega_\text{post}$& Rate of post-PEP protection waning (fixed at $1/14$) \\
\midrule
\multicolumn{3}{l}{\emph{Daily drivers (length \texttt{tf\_days}; scalars are broadcast)}}\\
\texttt{beta\_eff\_daily}  & $\beta_\text{eff}(t)$ & Daily transmission rate, inclusive of waves, seasonality, \\ && \quad and super-spreading \\
\texttt{f\_interv\_daily}  & $f_\text{interv}(t)$  & Daily contact-reduction multiplier $(1-c_\text{school})(1-c_\text{work})$,\\ && \quad  windowed \\
\texttt{importation\_daily}& $\lambda_\text{imp}(t)$ & Daily external importation intensity \\
\texttt{iso\_rate\_daily}  & $\nu_\text{iso}(t)$   & Daily isolation rate \\
\texttt{quar\_rate\_daily} & $\nu_\text{quar}(t)$  & Daily quarantine entry rate \\
\texttt{leave\_quar\_daily}& $\nu_\text{leave}(t)$ & Daily quarantine exit rate, no PrEP \\ && \quad ($0.05$ with vaccination, $0.10$ without) \\
\texttt{leave\_quar\_prep\_daily} & $\nu_\text{leave,prep}(t)$ & Daily quarantine exit rate, resume PrEP \\ && \quad ($0.05$ with vaccination, $0$ without) \\
\texttt{prep\_start\_daily}& $\rho_\text{start}(t)$& Daily PrEP/vaccination initiation rate; windowed \\
\texttt{prep\_stop\_daily} & $\rho_\text{stop}(t)$ & Daily PrEP discontinuation rate; not windowed \\
\texttt{pep\_rate\_daily}  & $\nu_\text{pep}(t)$   & Daily PEP administration rate \\
\texttt{prep\_eff\_daily}  & $\varepsilon_\text{prep}(t)$ & Daily PrEP/vaccine efficacy (susceptibility reduction)\\
\midrule
\multicolumn{3}{l}{\emph{Structure (Section \ref{sec:structure})}}\\
\texttt{n\_waning\_stages} & $n_w$                 & Number of sequential $R$ compartments, $1$--$3$ \\
\texttt{no\_latent}        & ---                   & Bypass $L$/$Q_L$; infection branches at transmission \\
\texttt{sis\_mode}         & ---                   & Retarget all recoveries to $S$ \\
\midrule
\multicolumn{3}{l}{\emph{Reactive intervention controller (Section \ref{sec:interv}); ignored when \texttt{reactive\_mode} $=0$}}\\
\texttt{reactive\_mode}    & ---                   & Enable threshold-triggered interventions \\
\texttt{on\_threshold}     & $\theta_\text{on}$    & Daily incidence at which interventions engage \\
\texttt{off\_threshold}    & $\theta_\text{off}$   & Daily incidence at which interventions release \\
\texttt{trigger\_delay}    & $\delta$              & Days between threshold crossing and activation \\
\texttt{reactive\_contact\_mult} & $c_\text{react}$& Multiplier on $\beta_\text{eff}(t)$ while active \\
\midrule
\multicolumn{3}{l}{\emph{Integrator (Section \ref{sec:tau})}}\\
\texttt{epsilon}           & $\epsilon$            & Leap error control (default $0.03$) \\
\texttt{Ncritical}         & $N_\text{crit}$       & Criticality threshold (default $10$) \\
\texttt{exactThreshold}    & $\Theta$              & SSA switch threshold (default $10$) \\
\texttt{maxtau}            & $\tau_\text{max}$     & Optional cap on the leap size (default $\infty$) \\
\bottomrule
\end{tabular}
\caption{Model parameters for the human-to-human simulator. Time-series parameters are vectors of length \texttt{tf\_days}; a single value is broadcast across all days.}
\label{tab:h2h-params}
\end{table}

\section{Vector-borne Simulator}
\label{sec:vector}

\subsection*{Compartments}

The vector-borne simulator extends the human compartment structure with a vector population in each spatial patch. For each patch $q\in\{1,\ldots,P\}$, four vector variables are tracked:
\[
S_{v,q},\qquad E_{v,q},\qquad I_{v,q},\qquad C_{v,q},
\]
representing susceptible, exposed, infectious, and cumulative infected vectors, respectively. The living vector population in patch $q$ is
\[
N_{v,q}=S_{v,q}+E_{v,q}+I_{v,q}.
\]
Vectors do not move between patches or have age structure. Human movement couples patches by determining the human population present and exposed to vectors in each patch. The total state-vector length is therefore $13PA+4P$.

\begin{table}[H]
\centering
\begin{tabular}{ll}
\toprule
Symbol & Description \\
\midrule
 $S_v$ & Susceptible vectors \\
$E_v$ & Exposed vectors (vector incubation) \\
$I_v$ & Infectious vectors \\
 $C_v$ & Cumulative vector infections \\
\bottomrule
\end{tabular}
\caption{Vector variables tracked in each of the $P$ spatial patches.
The total state-vector length is $13PA+4P$.}
\end{table}

\subsection*{Force of infection}

Define the biting-weighted human population present in patch $q$ as
\begin{equation}
D_q = \sum_{p=1}^{P} M_{pq} \sum_{a=1}^{A} b_a N_{p,a},
\label{eq:vector-effective-pop}
\end{equation}
where $M_{pq}$ is the fraction of time residents of patch $p$ spend in patch $q$. The age-specific biting weights are normalized so that their population-weighted mean is one.

For convenience, define

\begin{equation}
\beta_r(t) = \beta_{\mathrm{eff}}(t)f_{\mathrm{interv}}(t),
\end{equation}

where $\beta_{\mathrm{eff}}(t)$ already includes epidemic-wave, seasonal, and super-spreading multipliers.

\subsubsection*{Vector-to-human transmission}

The force of infection on humans in stratum $(p,a)$ is
\begin{equation}
\lambda^h_{p,a}(t) = \beta_r(t)\,b\,b_a\,\beta_{MH} \sum_{q=1}^{P} M_{pq}\frac{I_{v,q}}{D_q}.
\label{eq:vector-to-human-foi}
\end{equation}

\subsubsection*{Human-to-vector transmission}

The force of infection on susceptible vectors in patch $q$ is
\begin{equation}
\lambda_{v,q}(t) = \beta_r(t)\,b\,\beta_{HM} \frac{\displaystyle \sum_{p=1}^{P}M_{pq} \sum_{a=1}^{A} b_a\left(I_{s,p,a}+\alpha I_{a,p,a}\right)}{D_q}.
\label{eq:human-to-vector-foi}
\end{equation}

Here $b$ is the mean number of bites per vector per day, $\beta_{MH}$ and $\beta_{HM}$ are per-bite transmission probabilities, and $b_a$ represents age-specific relative exposure. Isolated infectious individuals do not contribute to transmission.

\subsection*{Transition Rates}

All human transitions from Section~\ref{sec:h2h} apply unchanged. The following vector reactions are added:

\subsubsection*{Vector Infection Cycle}
\begin{align}
  S_{v,q} \to E_{v,q}
    &\qquad \text{rate: } \lambda_{v,q}(t)\, S_{v,q} \\[4pt]
  E_{v,q} \to I_{v,q}
    &\qquad \text{rate: } \sigma_v\, E_{v,q} \\[4pt]
  \text{Vector Births} \to S_{v,q}
    &\qquad \text{rate: } \mu_v\, N_{v,q}
      \quad \text{(vector birth)} \\[4pt]
  S_{v,q} \to \text{Vector Deaths}
    &\qquad \text{rate: } \mu_v\, S_{v,q} \\[4pt]
  E_{v,q} \to \text{Vector Deaths}
    &\qquad \text{rate: } \mu_v\, E_{v,q} \\[4pt]
  I_{v,q} \to \text{Vector Deaths}
    &\qquad \text{rate: } \mu_v\, I_{v,q}
\end{align}

where $\sigma_v$ is the vector progression rate and $\mu_v$ is the vector mortality rate.

\subsection*{Parameters}

\begin{table}[H]
\centering
\begin{tabular}{lll}
\toprule
Parameter & Symbol & Description \\
\midrule
\texttt{sigma\_v} & $\sigma_v$ & Vector progression rate ($E_v \to I_v$) \\
\texttt{mu\_v}    & $\mu_v$    & Vector mortality rate \\
\texttt{a\_bite}  & $b$        & Mean biting rate (bites per vector per day) \\
\texttt{bite\_age\_weights} & $b_a$ & Age-specific relative biting exposure; normalized \\
                  &            & \quad so $\sum_a b_a N_{p,a} / \sum_a N_{p,a} = 1$ \\
\texttt{b\_h}     & $\beta_{MH}$ & Vector-to-human transmission probability per bite \\
\texttt{b\_v}     & $\beta_{HM}$ & Human-to-vector transmission probability per bite \\
\midrule
\texttt{f\_interv\_daily}  & $f_\text{interv}(t)$ & Daily contact-reduction multiplier, windowed \\
\texttt{Nv\_init\_frac}    & $m$        & Vector-to-human ratio $N_v/N_h$; scales transmission intensity \\
\texttt{Iv\_frac}          & ---        & Initial infectious fraction of the vector pool \\
\bottomrule
\end{tabular}
\caption{Vector-borne simulator additional parameters.}
\end{table}

\section{Environmental Simulator}
\label{sec:env}

\subsection*{Model Overview}

The environmental simulator extends the human-to-human model with a stochastic
environmental pathogen reservoir $W$. The human compartment structure is identical to Section~\ref{sec:h2h}, where $W$ is added as a single count after the 13 human blocks.  In particular, $W$ is measured in pathogen unit: one pathogen unit corresponds to $\kappa_W$ physical shedding units.

\subsection*{Additional State Space}

\begin{table}[H]
\centering
\begin{tabular}{lll}
\toprule
Symbol & Description \\
\midrule
 $W$ & Environmental pathogen concentration \\
\bottomrule
\end{tabular}
\caption{Environmental compartment. Total state vector length is
$13 \cdot P \cdot A + 1$.}
\end{table}

\subsection*{Force of Infection}

The force of infection on humans combines direct contact and waterborne
components:
\begin{equation}
  \lambda_{p,a}(t)
  = \underbrace{
      \beta_\text{eff}(t)\cdot f_c(t)
      \sum_{q=1}^{P} M_{pq}
      \sum_{b=1}^{A} C_{ab}
      \frac{I_{s,q,b} + \alpha\, I_{a,q,b}}{N_{q,b}}
    }_{\text{contact route}}
    \;+\;
    \underbrace{
      \delta(t)\cdot f_w(t)\cdot W(t)
    }_{\text{environment route}}
\end{equation}
where $\delta(t)$ is the environmental transmission coefficient (dose-response rate), $f_c(t)$ scales the contact route, and $f_w(t)$ scales the environment route.

\subsection*{Pathogen Dynamics}

\begin{align}
  \text{Shedding (symptomatic)} \to W
    &\qquad \text{rate: } \eta_I \sum_{p,a} I_{s,p,a} \\[4pt]
  \text{Shedding (asymptomatic)} \to W
    &\qquad \text{rate: } \eta_A\, \alpha_\text{env} \sum_{p,a} I_{a,p,a} \\[4pt]
  W \to \text{Decay}
    &\qquad \text{rate: } \mu_w\, W
\end{align}

\begin{equation}
  \frac{dW}{dt}
  = \underbrace{
      \eta_I \sum_{p,a} I_{s,p,a}
      + \eta_A\,\alpha_\text{env}\sum_{p,a} I_{a,p,a}
    }_{\text{shedding}}
    - \mu_w\, W
\end{equation}

where $\eta_I$ and $\eta_A$ are per-capita shedding rates from symptomatic and asymptomatic individuals, $\alpha_\text{env}$ is the relative environmental shedding of asymptomatic individuals, and $\mu_w$ is the pathogen decay rate.

\subsection*{Transition Rates}

All human transitions from Section \ref{sec:h2h} apply unchanged. 

\subsection*{Additional Parameters}

\begin{table}[H]
\centering
\begin{tabular}{lll}
\toprule
Parameter & Symbol & Description \\
\midrule
\texttt{eta\_I} & $\eta_I$ & Shedding rate from $I_s$ (packets per person per day) \\
\texttt{eta\_A} & $\eta_A$  & Shedding rate from $I_a$ (packets per person per day) \\
\texttt{alpha\_env}  & $\alpha_\text{env}$ & Relative asymptomatic environmental shedding \\
\texttt{mu\_w}  & $\mu_w$  & Pathogen decay rate in environment (day$^{-1}$) \\
\texttt{w\_unit} & $\kappa_W$   & Physical shedding units per pathogen unit; sets reservoir noise \\
\midrule
\texttt{delta\_eff\_daily} & $\delta(t)$  & Daily environmental transmission coefficient (per packet) \\
\texttt{f\_contact\_daily} & $f_c(t)$  & Daily contact-route scaling; carries scheduled closure \\
\texttt{f\_water\_daily} & $f_w(t)$ & Daily environment-route scaling ($\equiv 1$: closure does not \\
                           &             & \quad reduce environmental exposure) \\
\bottomrule
\end{tabular}
\caption{Waterborne simulator additional parameters.}
\end{table}

\section{Reproduction Numbers}
\label{sec:rt}

The effective reproduction number $R_t$ is computed once per simulated day by the next-generation matrix (NGM) method.

\subsection*{Shared infected subsystem}

The infected subsystem used to construct the next-generation matrix depends on the realized model structure for each simulation. We first describe the default structure with a latent stage. For every subpopulation--age cell $c=(p,a)$, the infected compartments are ordered
\begin{equation}
\bigl(L_c,; I_{s,c},; I_{a,c},; I_{iso,c},; Q_{L,c}\bigr).
\end{equation}
The NGM is built from two matrices indexed by (destination, source): $V$ collects transitions and removals within the infected subsystem, and $F$ collects the production of new infections. For structural variants, compartments and transition channels that are absent from the realized model are omitted or modified accordingly.

\paragraph{Transition matrix $V$.}
For the default latent-stage structure, $V$ is block-diagonal across cells. The per-cell block (rows and columns ordered
$L, I_s, I_a, I_{iso}, Q_L$) is

\begin{equation}
  V_c =
  \begin{pmatrix}
    \mu_L & 0 & 0 & 0 & 0 \\
    -\sigma(1-p_a) & \mu_{I_s} & -\kappa & 0 & 0 \\
    -\sigma p_a & 0 & \mu_{I_a} & 0 & -\sigma p_a \\
    0 & -\nu_\text{iso} & -\nu_\text{iso} & \mu_{I_{iso}} & -\sigma(1-p_a) \\
    -\nu_\text{quar} & 0 & 0 & 0 & \mu_{Q_L}
  \end{pmatrix},
\end{equation}
with diagonal removal rates (evaluated at day $t$)
\begin{align}
  \mu_L      &= \sigma + \nu_\text{quar}(t) + \nu_\text{pep}(t) + d, &
  \mu_{I_s}  &= \gamma + \nu_\text{iso}(t) + d, &
  \mu_{I_a}  &= \kappa + \nu_\text{iso}(t) + \gamma + d, \\
  \mu_{I_{iso}} &= \gamma + d, &
  \mu_{Q_L}  &= \sigma + \nu_\text{pep}(t) + d. &&
\end{align}

When the latent stage is absent, the $L$ and $Q_L$ states and their associated transitions are omitted from the infected subsystem. Other disabled transition channels give zero entries as appropriate in similar ways.

\paragraph{New-infection matrix $F$.}
New infections appear only in the latent rows $L_c$ when the latent stage is present. When the latent stage is absent, $L_c$ and $Q_{L,c}$ are omitted and the corresponding new-infections are instead split directly between the $I_s$ and $I_a$ rows with probabilities $1-p_a$ and $p_a$, respectively. Susceptibility is weighted by the effective susceptible count $\tilde S_{p,a} = S_{p,a} + (1 - \varepsilon_\text{prep}(t))\, S_{\text{prep},p,a}$. For the human-to-human model, the only nonzero entries link an infectious source cell $(q,b)$ to a latent destination cell $(p,a)$:
\begin{align}
  F\bigl[L_{p,a},\, I_{s,q,b}\bigr]
    &= \beta_\text{eff}(t)\, M_{pq}\, C_{ab}\,
       \frac{\tilde S_{p,a}}{N_{q,b}}, \\
  F\bigl[L_{p,a},\, I_{a,q,b}\bigr]
    &= \alpha\; \beta_\text{eff}(t)\, M_{pq}\, C_{ab}\,
       \frac{\tilde S_{p,a}}{N_{q,b}}.
\end{align}

\paragraph{Reproduction number.}
The effective reproduction number is the spectral radius of the NGM,
\begin{equation}
  R_t = \rho\bigl(F V^{-1}\bigr)
      = \max_i \mathrm{Re}\,\lambda_i\bigl(F V^{-1}\bigr).
  \label{eq:Rt}
\end{equation}

\subsection*{Vector-borne extension}

For a vector-borne simulation, the infected subsystem contains the human infected states described above together with $(E_{v,q},I_{v,q})$ for each patch $q$. Vector-borne simulations do not include direct human-to-human transmission. Therefore, the human-to-human entries of $F$ from the preceding subsection are set to zero and replaced by the following human-vector cross-transmission entries.

For every human stratum $(p,a)$ and vector patch $q$,
\begin{align}
F\bigl[L_{p,a},I_{v,q}\bigr] &= \beta_r(t)\,b\,b_a\,\beta_{MH}\,M_{pq} \frac{\widetilde S_{p,a}}{D_q},
\\
F\bigl[E_{v,q},I_{s,p,a}\bigr] &= \beta_r(t)\,b\,\beta_{HM}\,M_{pq}\,b_a \frac{S_{v,q}}{D_q},
\\
F\bigl[E_{v,q},I_{a,p,a}\bigr] &= \alpha\,\beta_r(t)\,b\,\beta_{HM}\,M_{pq}\,b_a \frac{S_{v,q}}{D_q},
\end{align}
where
\[
\widetilde S_{p,a} = S_{p,a} + \left(1-\varepsilon_{\mathrm{prep}}(t)\right) S_{\mathrm{prep},p,a}.
\]

For each patch, the vector transition block of $V$, ordered as $(E_{v,q},I_{v,q})$, is
\begin{equation}
V_{v,q} = \begin{pmatrix} \sigma_v+\mu_v & 0\\ -\sigma_v      & \mu_v
\end{pmatrix}.
\end{equation}

The effective reproduction number is
\begin{equation}
R_t=\rho\!\left(F(t)V(t)^{-1}\right).
\end{equation}
Thus, Reservoir uses the augmented human-vector next-generation-matrix convention. Under this convention, $R_t$ is the square root of the expected number of completed human-vector-human transmission cycles.

\subsection*{Environmental extension}
Because the environmental reservoir relaxes much faster than the epidemic in typical regimes, $W$ is collapsed to its quasi-steady state. A single infectious individual maintains a reservoir contribution of $\eta_I/\mu_w$ (symptomatic) or $\eta_A\,\alpha_\text{env}/\mu_w$ (asymptomatic). The waterborne route therefore contributes, for every source cell $(q,b)$ and destination cell $(p,a)$,
\begin{align}
  F\bigl[L_{p,a},\, I_{s,q,b}\bigr] &\mathrel{+}=
    \delta(t)\, f_w(t)\, \tilde S_{p,a}\, \frac{\eta_I}{\mu_w}, \\
  F\bigl[L_{p,a},\, I_{a,q,b}\bigr] &\mathrel{+}=
    \delta(t)\, f_w(t)\, \tilde S_{p,a}\, \frac{\eta_A\,\alpha_\text{env}}{\mu_w},
\end{align}
in addition to the contact entries (scaled by $f_c(t)$).

\section{Observation and Surveillance Model}
\label{sec:obs}

The simulator we describe above produces true latent incidence. However, to make our simulations realistic, we also have an observation model that maps the true, unobserved incidence to a noisy observed signal like we might see in real surveillance. Thus, two outputs are retained for each run: a true and an observed time series.

\subsection*{Reporting overdispersion}
\label{sec:obs-od}

Let $y_t = \sum_{p,a}$ (daily new infections) be the aggregate true incidence on day $t$. A single negative-binomial dispersion parameter is drawn once per run, $r \sim \text{LogNormal}(\log 50,\, 1.0)$, and reported counts are
\begin{equation}
  \tilde y_t \sim \text{NegBin}\!\Bigl(\text{size}=r,\;
                  p = \tfrac{r}{r + y_t}\Bigr),
  \qquad
  \mathbb{E}[\tilde y_t] = y_t,\;\;
  \mathrm{Var}[\tilde y_t] = y_t + \tfrac{y_t^2}{r}.
\end{equation}

\subsection*{Case reporting effects}
Four more augmentations are applied to the case series, each enabled independently (so that for some simulations, there may be reporting delays, for example, but in some there may not).

\begin{enumerate}
\item \textbf{Under-reporting.} The reporting fraction improves
      logistically from an initial to a final rate.
\item \textbf{Weekday effects.} Each day is multiplied by a clipped Gaussian factor with weekday-specific mean and SD (lower on weekends).
\item \textbf{Laboratory batch noise.} Cases are grouped into batches of size $\sim\!\text{Poisson}(100)$. Each batch receives an accuracy $\sim\!\mathcal{N}(1, 0.08^2)$, with rare ``bad'' batches, and a high-load processing penalty $\exp(-0.2\,\max(0,\ell_t - 0.8))$ where $\ell_t$ is the load relative to the 90th percentile.
\item \textbf{Reporting delays.} Cases are redistributed forward in time by a delay kernel whose maximum delay shrinks logistically over the outbreak.
\end{enumerate}

\subsection*{Symptomatic, hospitalizations and deaths}
Symptomatic incidence per age group is $y^{\text{symp}}_{a,t} = (1-p_a)\, y_{a,t}$, where $y_{a,t}$ aggregates true incidence across patches for age $a$. Hospitalizations and deaths are generated per age group with overdispersion and lags:
\begin{itemize}
\item \textbf{Hospitalizations.} With age-specific probability $h_a$, the daily
      admissions are $n^{\text{hosp}}_{a,t}\sim\min\bigl(\text{NegBin}(\mu =
      h_a y^{\text{symp}}_{a,t}),\, y^{\text{symp}}_{a,t}\bigr)$, each assigned
      an admission lag $\sim\text{Gamma}(\text{shape},\text{scale})$ with
      age-jittered parameters.
\item \textbf{Deaths.} The effective case-fatality ratio is
      $\text{CFR}^{\text{eff}}_{a,t} = \min(0.95,\; m_a\, c\, e^{r_d (t-1)/(T-1)}\,
      \xi_t)$, combining the baseline $m_a$, a per-run lognormal multiplier $c$,
      a temporal drift $r_d$, and an optional surge factor $\xi_t$. Deaths are a
      negative-binomial thinning of admissions, with an additional death lag
      (optionally bimodal) added to the admission lag.
\item \textbf{Occupancy.} Each admission draws a length of stay
      $\sim\text{Gamma}(2.5, \text{scale}=4)$ and increments daily occupancy
      over its stay.
\end{itemize}
The true hospitalization and death series are then passed through a lighter reporting model (logistic under-reporting and weekend suppression) to give the reported series.

\subsection*{Wastewater and syndromic signals}
\paragraph{Wastewater.}
Incidence is convolved with a normalized gamma shedding kernel over lags
$0,\dots,29$,
\begin{equation}
  k_d \propto \text{Gamma}\bigl(d;\, \text{shape}=\phi,\,
        \text{rate}=\phi/\bar d\bigr),
  \qquad
  W^{\text{raw}}_t = \sum_{d} k_d\, y_{t-d},
\end{equation}
with mean shedding delay $\bar d$ sampled per run. The signal is smoothed, multiplied by lognormal measurement noise, and normalized by a fecal-marker (PMMoV) series $\sim\!\text{pmmov\_mean}\cdot\text{LogNormal}(0,\sigma^2)$ to give a concentration analogue.

\paragraph{Syndromic surveillance.}
A syndromic (e.g., ILI-consultation) signal thins symptomatic incidence by a consultation probability, applies lognormal noise, and smooths: $\text{synd}_t = \text{smooth}\bigl(p_\text{cons}\, y^{\text{symp}}_t \cdot \text{LogNormal}(0,\sigma^2)\bigr)$.

\section{Initial Conditions}
\label{sec:init}

Initial conditions can be specified in the following ways: 1) seeding a proportion of the population as infectious (\texttt{seed\_frac}) and declaring whether the pathogen is novel (\texttt{novel}; under endemic pathogen conditions, users can directly specify the fraction of the population that begins immune \texttt{immune\_frac}. 2) defiingin explicity counts are compartments using (\texttt{n\_seed}).

\paragraph{Allocation across strata.}
The population is distributed across the $P \times A$ strata according to the
patch and age fractions. Immune individuals are allocated in proportion to each stratum's own population and under waning immunity conditions, the immune population is divided evenly across the chain.

\paragraph{Seeding.}
All initial infections are placed in  a single stratum, which represents the point of introduction. Within that stratum they are spread across the infected compartments in proportion to how long an individual spends in each ($L$, $I_s$, $I_a$)

\paragraph{Vector and Environmental Transmission}
For vectorborne transmission simulations, each patch's vector pool is sized with a vector-to-human ratio with an infectious fraction drawn per run. For waterborne transmission simulations, initial infections start at the following:
\begin{equation}
    W_0 = \frac{\eta_I \sum_{p,a} I_{s,p,a} + \eta_A\, \alpha_\text{env}  \sum_{p,a} I_{a,p,a}}{\mu_w}
\end{equation}

\section{Hierarchical Scenario Sampler}
\label{sec:sampler}

Datasets are generated by repeatedly drawing a scenario from a hierarchical prior, building the corresponding configuration, simulating, and keeping runs that exceed a minimum case count (5 for endemic, 10 otherwise). The sampler draws are organized into shared epidemiological parameters, a population structure, a scenario (forcing, interventions, demography), and mechanism-specific blocks. Tables \ref{tab:prior-epi}--\ref{tab:prior-mech} list the priors. $\mathcal{U}$ denotes uniform, $\log\mathcal{U}$ log-uniform, $\mathcal{LN}$ lognormal. Each retained run stores the sampled parameters, the configuration, the true incidence and daily $R_t$, both observation outputs (Section \ref{sec:obs}), and the full state trajectory. Runs are grouped by mechanism (\code{base}, \code{vector}, \code{waterborne}).

\begin{table}[H]
\centering
\begin{tabular}{llll}
\toprule
Quantity & Prior & Notes / clipping & Citation \\
\midrule
Infectious period (d) & $\mathcal{LN}(\log 5, 0.5)$  & clip $[1,30]$ & \cite{hakki2022onset, gani2004epidemiologic} \\
Latent period (d)     & $\mathcal{LN}(\log 5, 0.6)$  & clip $[0.5,21]$ & \cite{kiang2025modeling, xin2022estimating}  \\
Immunity duration (d) & $\mathcal{LN}(\log 365, 1.0)$ & clip $[14, 10950]$ & \cite{edridge2020seasonal, amanna2007duration} \\
$p_a$ (asymptomatic)  & $\text{Beta}(3,7)$           & mean $0.30$& \cite{sah2021asymptomatic, leung2015fraction} \\
$\alpha$ (asymp.\ trans.) & $\text{Beta}(2,5)$       & mean $0.29$& \cite{buitrago2022occurrence} \\
$\kappa$ ($I_a\!\to\!I_s$) & $\exp \mathcal{U}(\log 0.02, \log 0.5)$ && \\
$n_w$ (waning stages) & $\{1,1,1,2,3\}$ uniform pick & mostly $1$ & \cite{wearing2005appropriate} \\
\bottomrule
\end{tabular}
\caption{Shared epidemiological priors (\code{sample\_epi\_params}).}
\label{tab:prior-epi}
\end{table}

\begin{table}[H]
\centering
\begin{tabular}{llll}
\toprule
Mechanism & Quantity & Prior & Cite \\
\midrule
all          & $R_0$ (base)        & $\text{Gamma}(1.2,\,\text{rate}\,1.0)+0.8$ \;(mean $2.0$) & \cite{ke2021estimating, nikbakht2019comparison} \\
             & $R_0$ (vector)      & $\text{Gamma}(3.2,\,\text{rate}\,0.9)+0.4$ \;(mean $3.96$) & \cite{liu2020reviewing} \\
             & $R_0$ (environmental)  & $\text{Gamma}(2.6,\,\text{rate}\,0.85)+0.4$ \;(mean $3.46$) & \cite{mukandavire2011estimating}\\
\midrule
vector       & $\sigma_v$          & $\exp\mathcal{U}(\log\tfrac1{21}, \log\tfrac14)$ & \cite{liu2020reviewing} \\
             & $\mu_v$             & $\exp\mathcal{U}(\log\tfrac1{30}, \log\tfrac17)$  & \cite{liu2020reviewing} \\
             & $a$ (biting)    & $\mathcal{U}(0.1,1.0)$  &  \cite{liu2020reviewing}\\
             & $b_h, b_v$         & $\mathcal{U}(0.1,0.8)$ each &  \cite{liu2020reviewing} \\
             & $f_{Nv}, f_{Iv}$    &  $\mathcal{U}(0.05,2.0),\;\mathcal{U}(0.001,0.05)$ & \cite{liu2020reviewing}\\
\midrule
environmental   & environmental share    & $\mathcal{U}(0.1,0.9)$ of $R_0$ & \cite{mukandavire2011estimating} \\
             & $\eta_I$            & $\exp\mathcal{U}(\log 0.1, \log 1.0)$  & \cite{mukandavire2011estimating} \\
             & $\eta_A$            & $\exp\mathcal{U}(\log 0.05, \log 0.5)$ & \cite{mukandavire2011estimating} \\
             & $\alpha_\text{env}$ & $\mathcal{U}(0.1,1.0)$ & \cite{mukandavire2011estimating}  \\
             & $\mu_w$             & $\exp\mathcal{U}(\log 0.1, \log 0.5)$ & \cite{mukandavire2011estimating}  \\
\bottomrule
\end{tabular}
\caption{Mechanism-specific priors. All $R_0$ draws are floored at $0.3$. Transmission constants $\beta$ and $\delta$ are back-calculated from $R_0$.}
\label{tab:prior-mech}
\end{table}

\newpage

\begin{landscape}

\begin{table}[H]
\centering
\small
\begin{tabular}{llll}
\toprule
Quantity & Prior / construction & Notes & Citation \\
\midrule
$A$ (age groups) & pick from $\{1,1,1,2,3,5\}$ & $\Pr(A{=}1)=\tfrac12$ & \\
$P$ (patches)    & pick from $\{1,1,1,1,2,3\}$ & $\Pr(P{=}1)=\tfrac23$ & \\
Age fractions    & $\text{Dirichlet}$ via $\text{Gamma}(\alpha,1)$, $\alpha\!\sim\!\mathcal{U}(0.5,5)$ & & \\
Contact matrix $C$ & diagonal-dominant, age-distance decay, reciprocity-symmetrized & POLYMOD-like & \cite{kucharski2020effectiveness} \\
Mixing matrix $M_p$ & coupling $\sim\!\mathcal{U}(0.01,0.30)$, row-normalized & & \\
Structural mult.\ $\Phi$ & $\rho(M_0)$ & $=1$ if $P{=}A{=}1$ & \\
\midrule
Horizon \code{tf\_days} & fixed at $5,000$ for datasets & & \\
\#\,waves & $\{0,\dots,4\}$, $\Pr=(.35,.30,.20,.10,.05)$ & wave $\beta$: $\beta\cdot\mathcal{LN}(0,0.4)$, norm.\ & \\
Seasonality & enabled w.p.\ $0.8$; $1$--$3$ harmonics & periods $\{365,182.5,91.25\}$ & \\
 & amp.\ $\mathcal{U}(0.1,0.5)$, jitter $\mathcal{U}(0,30)$, daily noise $0.05$ & & \\
Super-spreading & $p_\text{ss}\!\sim\!\mathcal{U}(3\!\times\!10^{-4},10^{-2})$ & $\text{Gamma}(6,0.5)$ multiplier & \\
Intervention block & present w.p.\ $0.25$ & shared start/duration & \\
\;\; isolation & $\nu_\text{iso}\!\sim\!\mathcal{U}(0.05,0.5)$ & always, when block present & \cite{mossong2008polymod} \\
\;\; quarantine & $\nu_\text{quar}\!\sim\!\mathcal{U}(0.02,0.3)$ & w.p.\ $0.5$ & \cite{auranen2023efficacy} \\
\;\; PEP & $\nu_\text{pep}\!\sim\!\mathcal{U}(0.01,0.2)$ & w.p.\ $0.3$ & \cite{ambrosioni2021primary} \\
\;\; PrEP efficacy & $\varepsilon_\text{prep}\!\sim\!\mathcal{U}(0.5,0.95)$ & w.p.\ $0.2$ & \cite{arciuolo2017effectiveness} \\
Endemic regime & w.p.\ $0.8$ & enables births/deaths/importation & \\
\;\; birth rate $b$ & $\mathcal{U}(2\!\times\!10^{-5}, 1.2\!\times\!10^{-4})$ & $0$ if non-endemic & \\
\;\; death rate $d$ & $b\cdot\mathcal{U}(0.8,1.3)$ & $0$ if non-endemic & \\
\;\; importation $\lambda_\text{imp}$ & $\exp\mathcal{U}(\log 0.01, \log 0.5)$ & $0$ if non-endemic & \\
Seeding compartment & pick from $\{I_s,I_s,I_s,L,I_a\}$ & & \\
\#\,seeds $n_\text{seed}$ & $\mathcal{U}\{1,\dots,50\}$ & & \\
Population $N$ & $\exp\mathcal{U}(\log 5\!\times\!10^3, \log 4\!\times\!10^7)$ & log-uniform & \\
$h$ (hosp.\ prob) & $\mathcal{U}(0.005,0.15)$ & age-scaled & \\
$m$ (death prob)  & $\mathcal{U}(0.001,0.10)$ & age-scaled & \\
\bottomrule
\end{tabular}

\caption{Population-structure and scenario priors
(\code{sample\_population\_structure}, \code{sample\_scenario}).}
\label{tab:prior-scenario}
\end{table}

\end{landscape}

\end{appendices}

\end{document}